# Redox Active Cerium Oxide Nanoparticles: Current Status and Burning Issues

**Megan S. Lord[#$], Jean Francois Berret[#&], Sanjay Singh[#§], Ajayan Vinu[¤] and Ajay S. Karakoti*[¤]**

[$]Graduate School of Biomedical Engineering, UNSW Sydney, Sydney, New South Wales, Australia

[&]Matière et systèmes complexes, Université de Paris, CNRS, 75013 Paris, France

[§]Division of Biological and Life Sciences, Ahmedabad University, Ahmedabad, Gujarat, India

[¤]Global Innovative Center for Advanced Nanomaterials, College of Engineering, Science and Environment, The University of Newcastle, Callaghan, New South Wales, Australia

[#]Authors have equal contribution

*Corresponding author

**Abstract**

Research on cerium oxide nanoparticles (nanoceria) has captivated the scientific community due to their unique physical and chemical properties, such as redox activity and oxygen buffering capacity, which made them available for many technical applications, including biomedical applications. The redox mimetic antioxidant properties of nanoceria have been effective in the treatment of many diseases caused by reactive oxygen species (ROS) and reactive nitrogen species. The mechanism of ROS scavenging activity of nanoceria is still elusive, and its redox activity is controversial due to mixed reports in the literature showing pro-oxidant and antioxidant activity. In lieu of its current research interest, it is critical to understand the behaviour of nanoceria in the biological environment and provide answers to some of the critical and open issues. This review critically analyses the status of research on the application of nanoceria to treat diseases caused by ROS. It reviews the proposed mechanism of action and shows the effect of surface coatings on its redox activity. It also discusses some of the crucial issues in deciphering the mechanism and redox activity of nanoceria and suggests areas of future research.

## 1.0 Introduction

Redox homeostasis is essential for normal tissue function with a low level of reactive species required for normal cell signalling. These reactive species include oxygen, nitrogen, sulfur and chloride species, with oxygen radicals being the most abundant. The major drivers of various reactive oxygen species (ROS) and reactive nitrogen species (RNS) produced in the cells are superoxide radicals ($O_2^-$), hydrogen peroxide ($H_2O_2$), hydroxyl radicals (•OH), peroxynitrite ($ONOO^-$) and nitric oxide (NO). In a state of redox homeostatasis, cells have the ability to neutralize ROS with the antioxidant systems such as the enzymes glutathione peroxidase (GPx), catalase (CAT) and superoxide dismutase (SOD). However, when redox homeostasis is imbalanced, excessive ROS can lead to oxidative stress, which irreversibly damages proteins, lipids and DNA that underlie the pathogenesis of many chronic conditions such as diabetes, cardiovascular diseases, neurodegenerative disorders and cancer (**Figure 1**)[1]. Antioxidants neutralize free radicals to re-establish redox homeostasis and modulate ROS-mediated cell signalling pathways. The human body produces multiple intracellular and extracellular antioxidants to neutralize the free radicals, and several other naturally existing antioxidants can be supplemented through the human diet (**Table 1**).

The recent emergence of nanotechnology has enabled the development of nanoparticles (NPs) that can also serve the same purpose of acting as naturally existing enzyme involved in production or scavenging of free radicals. These nanomaterials are often called as nanozymes owing to their ability to mimic the activity of enzymes. Even though the term nanozyme has stirred up some debate in the literature, we will use this term, where necessary, in this review to be consistent with the previous results and terminology[2]. Among different nanoparticles depicting enzyme mimicking properties, cerium oxide nanoparticles (nanoceria), a rare earth metal oxide having a fluorite structure, have been receiving a lot of attention owing to their ability to mimic the activities of antioxidant enzymes, such as SOD and CAT. As nanoceria exhibits unique redox catalytic properties and a high chemical and thermal stability, it has traditionally been employed as a catalyst in its bulk and nano-size forms[3] in water gas shift catalysis[3a], CO oxidation[4], catalytic convertor in the automobiles[5] and methanol fuel cells[6]. It is also used as an oxygen buffer due to its ability to reversibly gain or release oxygen molecules (or transport oxygen anions ions) at high temperature while keeping its fluorite structure intact[7]. This activity is highlighted by the presence of surface and sub-surface oxygen vacancies in nanoceria and has been used in designing oxygen sensors, oxygen ion-conducting membranes[8] (for use in solid oxide fuel cells)[9] as well as thermo-catalytic production of hydrogen from water[10]. Cerium oxide is also used as an abrasive and polishing material in the chemical and mechanical planarization of silicon wafers[11] wherein, higher $Ce^{3+}$ content is associated with higher material removal rates[12]. Thus, it is evident that the technological importance of nanoceria stems from its redox capability and oxygen buffering capacity.

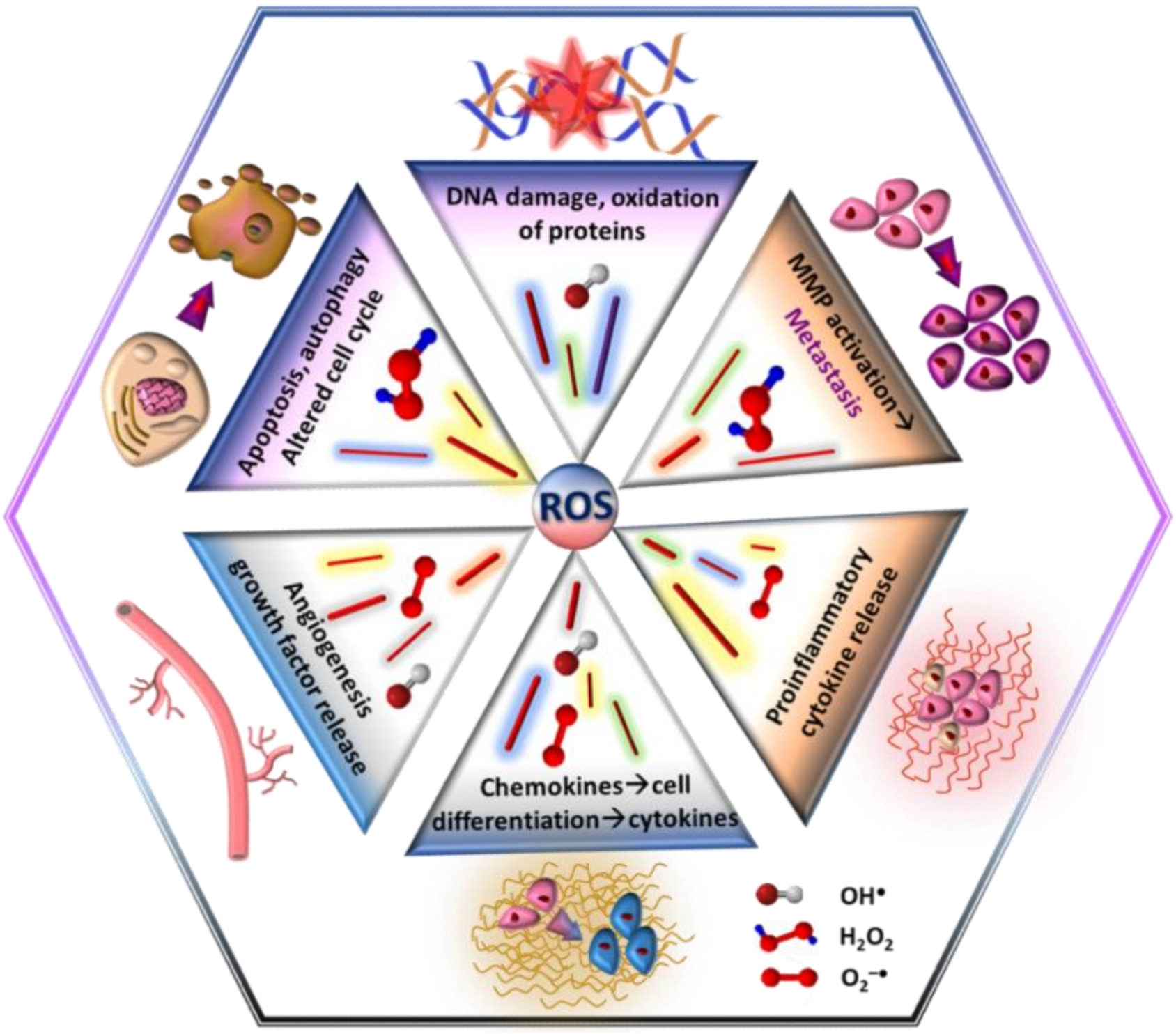

***Figure 1.*** *Schematic of the involvement of ROS in many biological processes suggesting the role of endogenous and exogenous ROS in regulating cellular processes.*

These unique properties exhibited by nanoceria make it available for potential therapeutic application for an array of biological conditions that exhibit oxidative stress. Nanoceria act as biological antioxidants due to the surface Ce(III)/Ce(IV) valency switch that provides ROS scavenging properties[13]. This is clearly reflected by the fact that the redox activity and oxygen buffering capacity of nanoceria have been explored extensively for various biomedical applications[14]. For example, nanoceria interacts with a range of ROS and RNS species as the redox potential of nanoceria is uniquely positioned to interface with ROS and reduce superoxide radicals or oxidize hydrogen peroxide[15] and has been shown to alleviate oxidative stress in a multitude of disease models. The small energy difference between its two oxidation states, $Ce^{3+}$ and $Ce^{4+}$, allows for easy cycling based on subtle differences in the redox environment.

**Table 1: Naturally existing antioxidants for scavenging ROS and RNS and their mechanisms**

| Natural Enzyme | Cellular location | Mol. Weight (kDa) | Reaction mechanism | Free radicals generated/ scavenged | Major functions in cells/tissues | Ref |
|---|---|---|---|---|---|---|
| **Cytoplasmic Cu/Zn Superoxide dismutase (SOD1)** | Cytoplasm, mitochondria intermembrane space, nucleus, lysosome, peroxisomes | 32 | $Cu^{2+}ZnSOD + O_2^{\bullet -} \rightarrow Cu^{+}ZnSOD + O_2$<br><br>$Cu^{+}ZnSOD + O_2^{\bullet -} + 2H^{+} \rightarrow Cu^{2+}ZnSOD + H_2O_2$ | $O_2^{\bullet -}$ and $H_2O_2$ generation | RNA metabolism as nuclear transcription factor, Regulating metabolic signaling by converting $O_2^{\bullet -}$ into $O_2$ and $H_2O_2$. | [16] |
| **Mitochondrial MnSOD (SOD2)** | Mitochondria matrix | 96 | $Mn^{3+}SOD + O_2^{\bullet -} \rightarrow Mn^{2+}SOD + O_2$<br><br>$Mn^{2+}SOD + O_2^{\bullet -} + 2H^{+} \rightarrow Mn^{3+}SOD + H_2O_2$ | $O_2^{\bullet -}$ conversion to $H_2O_2$ | Plays vital role in mitochondrial death pathway, cardiac myocyte apoptosis signaling | [17] |
| **Extracellular Cu/Zn SOD (SOD3, ecSOD)** | Extracellular matrix, cell surface, extracellular fluid | 135 | $Cu^{2+}ZnSOD + O_2^{\bullet -} \rightarrow Cu^{+}ZnSOD + O_2$<br><br>$Cu^{+}ZnSOD + O_2^{\bullet -} + 2H^{+} \rightarrow Cu^{2+}ZnSOD + H_2O_2$ | $O_2^{\bullet -}$ and $NO^{\bullet}$ scavenging. $H_2O_2$ production | Prevents superoxide-mediated oxidation of $NO^{\bullet}$ and controls the formation of $ONOO^{-}$, defense mechanism in lungs, Protection from exogenous stress | [18] |
| **CAT** | Peroxisomes and cytosol of erythrocytes | Small Subunit: 55-69<br><br>Large Subunit: 75-84 | $Fe^{3+}$-Porphyrin + $H_2O_2 \rightarrow O{=}Fe^{4+}$-Porphyrin + $H_2O$<br><br>$O{=}Fe^{4+}$-Porphyrin + $H_2O_2 \rightarrow Fe^{3+}$-Porphyrin + $H_2O$ | Decomposition of $H_2O_2$ | Protects haemoglobin and pancreatic β-cells from oxidative stress, utilizes $H_2O_2$ during the process of fatty acid oxidation | [19] |
| **Alkaline Phosphatase (ALP)** | Plasma membrane of liver, bone, kidney, and intestinal mucosa, bile duct, and placenta | 86 | $ALP + R_1OP \rightarrow ALP\bullet R_1OP \rightarrow ALP\text{-}P + R_1OH$<br><br>$ALP\text{-}P + H_2O \rightarrow ALP\bullet P \rightarrow ALP + P_i$<br><br>$ALP\text{-}P + R_2OH \rightarrow ALP\bullet R_2OP \rightarrow ALP + R_2OP$ | Phosphate removal | Bone mineralization, and break-down of proteins, transportation of various biomolecules (fatty acid, phosphates and calcium) in the intestine, and regulation of cell growth during fetal development | [20] |
| **Peroxidase** | Lysosomes, cytoplasm, exocrine secretion (tears, saliva, vaginal fluid) | Glutathione peroxidase: 83-95, | $H_2O_2 + AH_2 +$ Peroxidase $\rightarrow 2H_2O + A +$ Peroxidase<br>$H_2O_2 + 2AH_2 +$ Peroxidase $\rightarrow$ | Decompose $H_2O_2$ and $OH^{\bullet}$ generation | Plays major role in innate immunity and regulation of apoptosis or cell signaling process | [21] |
| | | Thyroid peroxidase: ~100,<br><br>Salivary peroxidase: 78-80 | $2H_2O + 2AH^{\bullet}$ + Peroxidase | | | |
| **Oxidase** | Mitochondrial respiratory chain | Glucose oxidase: 160, Xanthine oxidase: 275, Cytochrome c oxidase: 200 | $O_2$ + AH + Oxidase → $H_2O$ + A + Oxidase<br><br>$H_2O$ + AH + Oxidase + $O_2$ → $H_2O_2$ + A + Oxidase | $H_2O_2$ generation | Helps in respiratory chain by reducing the molecular oxygen to water, convert glucose into gluconic acid, and catalyze the oxidation of xanthine to uric acid. | [22] |
| **Glutathione (GSH)** | Liver cytosol, extracellular space | 307.32 Da | GSH + $^{\bullet}$R → $GSH^{\bullet+}$ + $R^-$<br>GSH + $^{\bullet}$R → GSH(-H)$^{\bullet}$ + HR | OH$^{\bullet}$ scavenging | Reduce oxidative stress, fight autoimmune diseases, and reduce cell damage in fatty liver diseases | [23] |
| **Vitamin E (α Tocopherol)** | Liver and fatty tissue | 430.71 Da | LOO$^{\bullet}$ + α Tocopherol-OH → LOOH + α Tocopherol-O$^{\bullet}$ | OH$^{\bullet}$ scavenging | Inhibits the free radical production during the oxidation of fats | [24] |

LOO$^{\bullet}$ - Lipid peroxyl radical, $^{\bullet}$R – peroxyl radical, A-substrate.

Nanoceria's ability to scavenge free radicals and act as antioxidant has been tested in many applications such as protection of eye from macular degeneration, pulmonary radioprotection during radiotherapy[25], liver protection[26] from ROS, reduction in inflammation in Inflammatory Bowel Disease (IBD)[27] and protect against AB induced neuronal cell death[28]. While the nanoceria's antioxidant abilities are well documented, it has also been shown that the redox property of nanoceria can be used for oxidizing various substrates at acidic pH values to generate a colorimetric response for the development of sensors[29]. This ability to oxidize substrates such as tetramethylbenzidine (TMB), azino-bis(3-ethylbenzothiazoline-6-sulfonic acid) ABTS, o-phenylenediamine (OPD) and dopamine have been used for creating pH or redox responsive sensors for cancer detection[30], dopamine sensing[31] and even generate oxidative stress for tumour therapy[32]. These oxidative properties (oxidase-like or peroxidase-like) opened a plethora of opportunities for the biomedical applications of nanoceria[33], dividing the scientific community between its pro-oxidant and antioxidant properties. This is especially important for future biomedical applications of nanoceria wherein only anti-oxidative properties are required. Owing to these conficiting anti-oxidative and oxidative properties of nanoceria and its protective as well as potential toxic effect on cells respectively, the research and translation of nanoceria into nanotherapeutics is severely hampered. Similarly, the scientific community is uncertain about the effect of oxidation state and other important physico-chemical properties of nanoceria on its oxidative potential that is very important for its translation to real world sensors. Even though a few reviews exist on therapeutic potential of redox responsive nanomaterials[34] for therapeutic applications and specifically on synthesis and applications of nanoceria[13a, 13c, 35], these reviews only cover the conflicting reports on nanoceria without critical analysis of the physico-chemical parameters that are responsible for such conflicting observations. Some of the reviews are focused on the physical properties of nanoceria without taking into account the biological applications[13c, 35b] while almost none of the reviews address the role of surface coating on tuning its activity or only partially addresses it[36]. Thus, this review focuses on describing various enzyme–mimicking activities of nanoceria and divulging the currently proposed mechanisms of activity. It describes the current applications of nanoceria across disease models along with the role of intentional biocompatible surface coatings on the activity of nanoceria. Finally, it reviews the current issues in deciphering the mechanism of action of nanoceria, specifically emphasizing the role of vacancies and oxidation state and its critical anti-oxidative versus oxidative potential. The review concludes with recommendations for future research that could potentially answer some of the critical challenges in nanoceria research and propel the science towards a fruitful commercial outcomes.

### 1.1 Unique features of cerium oxide as compared to other nanozymes (role of size, morphology and oxidation state)

In 2005, it was shown that the vacancy engineered nanoceria could confer protection from radiation-induced cellular damage[25]. This was the first report depicting the use of an inorganic metal-based nanomaterial to prevent ROS-mediated damage in a biological system. Prior to this, only fullerenes[37], bare and water-soluble forms, were shown to possess an enzyme-like antioxidant activity, while the majority of other material systems at that time was considered pro-oxidant[13b, 38]. Post this report, selected inorganic systems including yttrium oxide[39], vanadium oxide[40], manganese oxide[41], cobalt oxide[42], iron oxide[43], gold[44] and platinum[45] were also reported to show antioxidant activity, specifically scavenging hydrogen peroxide and superoxide species[13b]. A few isolated reports of anti-oxidant activity of carbon-based materials such as carbon nanotubes, carbon nanoclusters, and polymer or inorganic materials coated carbon nanostructures have gained momentum recently[46]. All these materials have shown antioxidant behavior mimicking the SOD-, CAT- or GPx-like activities. In this sense, most nanomaterials with antioxidant activity behave like oxido-reductase enzymes and contribute to the antioxidant mechanism through oxidation or ROS reduction.

Unlike most metals and metal oxides that typically demonstrate a size-dependent or an oxidation state dependent enzyme activity, nanoceria is unique in depicting a combined size-and oxidation state-dependent activity. For example, popular nanozymes like $V_2O_5$ nanowires, $Mn_xO_y$ and $Co_3O_4$ show their enzyme mimicking activities dependent on the active oxidation state of the metal ions[40-42]. These activities are unidirectional i.e. the oxidation state can change only in one direction resulting in either the oxidation or reduction of metal ions upon interaction with ROS/RNS. Metallic NPs such as Au, Ag and Pt show catalytic activity owing to the ability of the metal ions to act as catalytically active centers for adsorption of oxygen radicals followed by an electron transfer[38, 45]. The reduction in the size of these nanomaterials results in increasing the total surface area of the materials, exposing a large number of active sites. These active sites can function as catalytic centers for reduction or oxidation reactions, and hence the catalytic activity increases with the reduction in the size of these nanomaterials. However, the overall mechanism of imparting several different types of activities is unclear for some of these materials. For example, Pt NPs demonstrate both CAT- and SOD-like activities at near neutral pH [45, 47]. Similarly, $Mn_xO_y$ and $V_2O_5$ demonstrate both peroxidase-like and GPX-like activity with similar compositions[40-41], and it is unclear if different catalytic activities shown by these particles follow a different mechanism pathway that is dependent on the size, charge or oxidation state of the NPs.

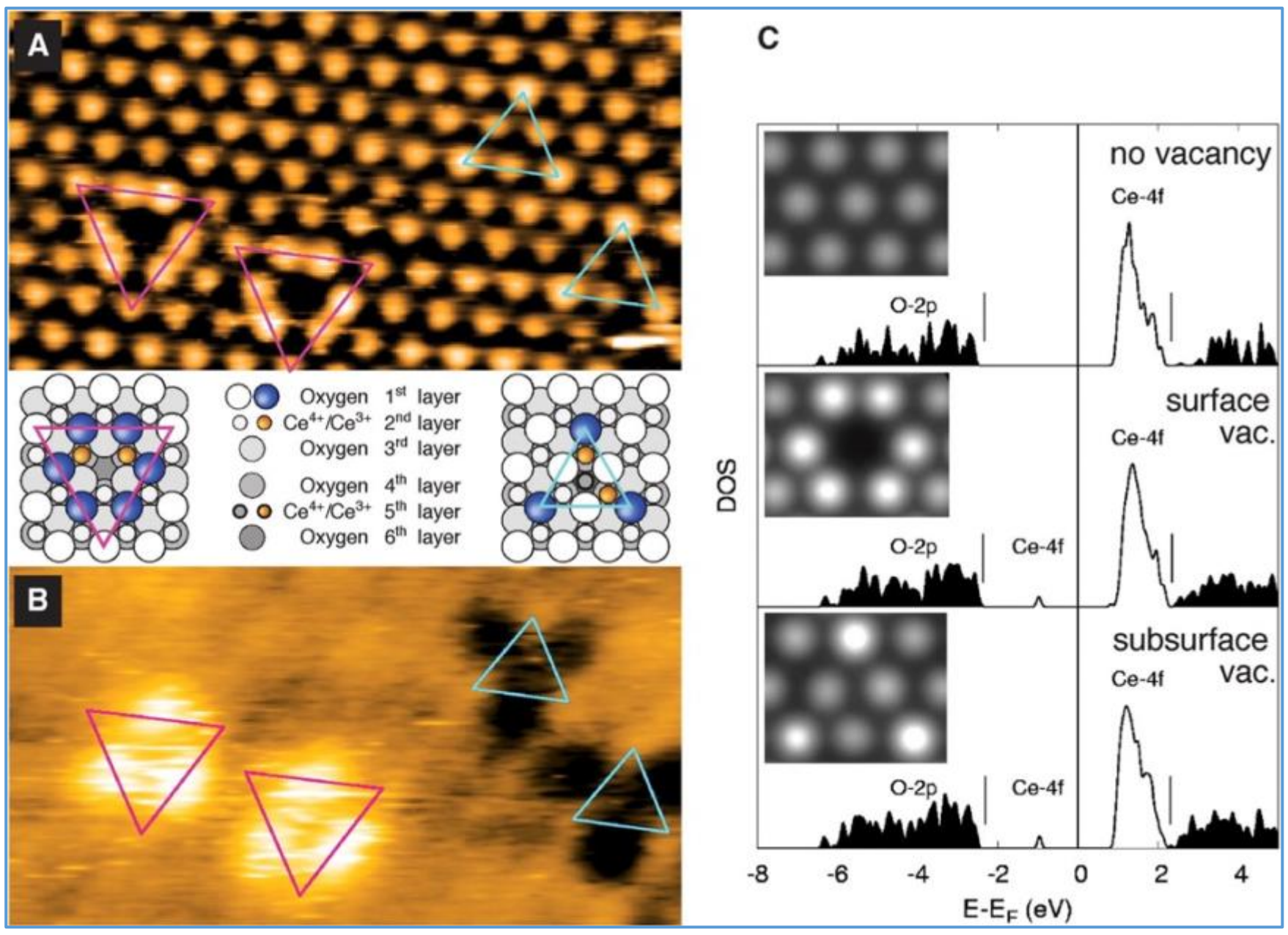

***Figure 2***: *(**A**) Filled-state and (**B**) empty-state STM images of single vacancies and related structural models (left, surface vacancy; right, subsurface vacancy; characteristic O rim atoms in blue). (**C**) Calculated density of states (DOS) and simulated filled-state STM images (bias -3.0 V). Ce 4f gap states, displayed as unshaded curves, do not contribute to the STM images because of their strong spatial localization. STM imaging conditions: 6.6 nm by 3.5 nm; -3.0 V (A), +3.0 V (B); 0.1 nA (Reproduced with permission from 38).*

Compared to other nanozymes, several mechanistic studies have been carried out to understand the antioxidant behavior of nanoceria and understand the influence of size and oxidation state on its activity[48]. It is well known in the literature that reducing the size of nanoceria results in the creation of surface oxygen vacancy to compensate for the increased surface energy[49]. The leaving oxygen molecule donates two electrons that localise in the *f* orbital, causing the reduction of two Ce(IV) cations to Ce(III) cations while preserving the overall fluorite lattice of the cerium oxide[7]. The single oxygen vacancies have been characterized as either surface or sub-surface vacancies and differ in the electronic and structural arrangement as shown in **Figure 2[50]**. It has been argued that the sub-surface oxygen vacancies are more stable than the surface oxygen vacancies, while the electrons left by oxygen do not localize on the nearest cerium ions close to vacancies but on the next nearest cerium neighbours[51]. This transition is more pronounced for smaller sized NPs with high surface energy than large-sized particles[52]. Thus, changing the size alters the ratio (Ce(III)/Ce(IV)) of the surface oxidation states, and hence nanoceria shows both size and oxidation state dependent activity.

Interestingly, surface crystal facets and synthesis methods also play a role in creating size-induced oxygen vacancies, thereby controlling the oxidation state of surface cerium ions in nanoceria[53]. This allows the oxidation state of nanoceria to be controlled based on its surface composition from $CeO_2$ to $CeO_{2-x,}$ with x being as high as 0.2[54]. Nanoceria also shows oxidation state-dependent enzyme mimicking activity wherein it catalyses the reduction of superoxide radicals at higher surface Ce(III)/Ce(IV) concentration[48b] while it shows CAT activity (i.e decomposition of hydrogen peroxide) with negligible SOD activity at lower Ce(III)/Ce(IV) concentrations[48c]. Thus its, enzymatic activity is not unidirectional and can result in both reduction and oxidation of surface cerium ions depending upon the pH and the redox potential of the species involved. Similarly, the morphology of nanoceria is expected to play a role in its antioxidant or pro-oxidant activity as the different morphologies present different crystalline facets for reaction with ROS[55]. Nanoceria with the rod, cube and spherical morphologies show (110), (100) and (111) type crystal facets with facet-dependent redox properties, which play a key role in controlling the peroxidase and SOD activity[56]. For example, nanoceria with cube morphology and exposed (100) facets offer a higher peroxidase but lower SOD activity as compared to that of nanoceria with rod-shaped morphology. It must be noted that parameters such as the shape, the route of synthesis and the surface charge are relevant in controlling either antioxidant or pro-oxidant activities of nanoceria only when these parameters can affect the $Ce^{3+}/Ce^{4+}$ ratio on its surface.

The mechanism of SOD- and CAT-like activity of nanoceria is also different from most reported nanomaterials. For example, in SOD-like activity, nanoceria reduces superoxide radicals at neutral pH to hydrogen peroxide, although it cannot carry out the other half of the reaction i.e. oxidation of superoxide directly to oxygen[57]. While the redox potential of Ce(III)/Ce(IV) at neutral pH is sufficient to reduce superoxide radicals to peroxide, the reduction of surface Ce(IV) to Ce(III) by superoxide radicals is not feasible on the surface of nanoceria as supported by some of the theoretical studies on the adsorption of oxygen on surface vacancies[58]. This behavior is notably different from the actual SOD enzyme and nanomaterials such as fullerenes that can catalyze both reduction and oxidation of superoxide radicals. Similarly, the CAT-like activity of nanoceria also follows a unique pathway. Nanoceria, depending upon its size and oxidation state, forms a unique cerium-oxo-peroxo complex with hydrogen peroxide that degrades over 7 – 21 days at room temperature[59]. This complex is stable for a longer time for smaller NPs with a higher Ce(III)/Ce(IV) ratio while it quickly dissociates to regenerate cerium oxide and oxygen for larger particles[52, 59a]. If we combine the ability of nanoceria to reduce superoxide radicals and oxidize peroxide ions to oxygen, a cyclic behavior is obtained wherein nanoceria can reversibly regenerate its oxidation state by interacting sequentially with superoxide and peroxide species. This gives nanoceria its famous regenerative capacity wherein a single dose of nanoceria, at a lower concentration, is expected to be effective in mitigating ROS-induced damage.

Nanoceria possesses oxidase-like activity that is also associated with its oxidative damage causing potential. While the oxidase-like activity is well documented, this activity is observed only at acidic pH[60]. In a series of reports, researchers have tried to establish the relationship between the redox potential of biologically active processes with the band gap of the material to determine their oxidative potential. Burello and Worth and Nel and colleagues demonstrated the potential of six metal oxide NPs whose band structure overlaps with the cellular redox potential and thus, can show harmful oxidative properties[61]. Even though nanoceria was not included in these reports, a later report by Liu et al, revealed nanoceria, along with many other metal oxides, was non-toxic based on QSAR models and cytotoxicity testing on two different cell lines[62]. Recently, Noventa et al. tested the band gap and the ion dissolution theory for the oxidative potential of nanoceria and found that nanoceria does not show any oxidative effects based on these theories, while ZnO and $MnO_2$ showed a dependence on the ion dissolution theory[63]. Thus, even though there are some documented reports on the oxidative and cytotoxic potential of nanoceria, it does not follow the QSAR-based theories of cytotoxicity applicable for other oxides and its oxidative potential is likely linked to its oxidase-like activity at acidic pH or due to the physical damage caused by the sharp faceted nanoparticles. The differences in cytotoxicity were also documented that sheds some light on the activity of nanoceria[35b]. Based on this discussion, it is clear that nanoceria behaves differently in biological media as compared to most other oxide-based nanomaterials due to its ability to interact reversibly in a pH-dependent manner with intracellularly generated ROS.

## 2.0 Proposed mechanism of action of catalytic activities displayed by nanoceria in cell-free and cell-based systems (Antioxidant and Pro-oxidant activities)

As discussed above, a plethora of studies indicate that the two oxidation states of the surface "Ce" ions of nanoceria Ce(III)/Ce(IV) predominantly control the observed catalytic activity[26, 34a]. The easy switchability between Ce(III) and Ce(IV) oxidation states leads to the regeneration of the required oxidation state, facilitating the continuous antioxidant catalytic activity of nanoceria. Blockage of this redox cycling leads to pro-oxidant catalytic activity, especially in acidic medium. The following section describes the antioxidant and pro-oxidant activities of nanoceria in detail.

**2.1 Antioxidant catalysis by nanoceria**: Nanoceria displays the catalytic activities analogous to several biological enzymes that are intrinsic constituents of the antioxidant cascade of mammalian cells/tissues. The key enzymes involved in the regulation of redox balance in the cytoplasm are frequently found to control cellular ROS and RNS. Most organisms succumb to the oxidative and nitrosative stress induced by $O_2^-$, $H_2O_2$, •OH, $ONOO^-$ and NO species. Mammalian cells have well-orchestrated defence mechanisms from ROS and RNS that constitute enzymatic radical scavengers, such as SOD, CAT, GPH, thioredoxin, peroxiredoxin, and glutathione transferase, and non-enzymatic scavengers, such as ascorbic acid, α-tocopherol, all trans retinol-2, and β-carotene, and glutathione[64].

**2.1.1 SOD-like activity**: In nature, three major families of SOD enzymes have been identified, which vary in protein folding and the embedded metal cofactor. These are Cu/Zn-SOD, Fe and Mn-SOD, and the Ni-SOD[65]. The $H_2O_2$ disproportion reaction catalyzed by these enzymes can be represented by a general equation –

$$M^{(n+1)+}\text{-SOD} + O_2^- \rightarrow M^{n+}\text{-SOD} + O_2$$
$$M^{n+}\text{-SOD} + O_2^- + 2H^+ \rightarrow M^{(n+1)+}\text{-SOD} + H_2O_2$$

Where, M = Cu (n = 1), Mn (n = 2), Fe (n = 2), Ni (n = 2), Zn (n = 2).
In the above two reactions, the oxidation state of the metal center switches between n and n+1. Analogous to this, the surface atoms of nanoceria also oscillate their oxidation state between Ce(III) and Ce(IV) and thus display SOD mimetic catalytic activity by following the series of reactions:

$$O_2^- + Ce^{+3} \rightarrow O_2^{2-} + Ce^{+4}$$
$$O_2^{2-} + 2Ce^{4+} \rightarrow O_2 + 2Ce^{3+}$$

Self and co-workers were the first to report SOD mimetic activity of nanoceria and proposed with substantial evidence that the oscillation of Ce(III) and Ce(IV) oxidation states of "Ce" as the possible mechanism[57]. They showed excellent activity in nanoceria (3–5 nm), with a catalytic rate ($3.6 \times 10^9$ $M^{-1}$ $s^{-1}$) better than the natural SOD enzyme (1.3 and $2.8 \times 10^9$ $M^{-1}$ $s^{-1}$). Subsequently, Celardo et al. provided a more comprehensive mechanism by a model of the sequence of reactions involved in the dismutation of superoxide anions by nanoceria[66], as shown in **Figure 3A**. The superoxide anion binds with the oxygen vacancies created around three Ce(III) ions in the crystal lattice of nanoceria. This binding facilitates an electron transfer from a Ce(III) atom to an oxygen atom and binding of two protons with the two electronegative oxygen atoms to form hydrogen peroxide. During this process, the Ce(III) ion is oxidized to Ce(IV) under the oxidizing influence of hydrogen peroxide. Interestingly, hydrogen peroxide also acts as a reducing agent and converts Ce(IV) ions into Ce(III), ready to bind with another superoxide anion and dismutate them[15] owing to the subtle changes in the environment and the fenton-like chemistry of Ce ions with hydrogen peroxide. Thus, under the influence of hydrogen peroxide, the nanoceria surface is auto-regenerated to facilitate the continuous scavenging of superoxide anions.

**2.1.2 Catalase-like activity**: The reversible reaction of soluble cerium ions with hydrogen peroxide at acidic pH has been studied in detail in the early 1950s. Pioneering work by Baer and Stein, Sigler and Masters and Bielski and Saito demonstrated a two-step redox activity where the reaction starts with the interaction between hydrogen peroxide and Ce(IV) and produces Ce(III), protons, water and molecular oxygen[67]. However, the CAT mimetic activity of nanoceria was first reported by Self and co-workers by using three independent methods, Amplex Red assay, the absorbance of hydrogen peroxide at 240 nm, and molecular oxygen probe[48c]. The results substantiated that nanoceria with Ce(IV) oxidation state atoms display better CAT-like activity than Ce(III). The breakdown of hydrogen peroxide in water and molecular oxygen was confirmed by an oxygen electrode that showed a concentration-dependent response. Celardo et al. further studied the mechanisms and hypothesized that hydrogen peroxide dismutation occurs via a two-step process (**Figure 3B**) [66]. In the first step, the interaction of one molecule of hydrogen peroxide with nanoceria leads to the reduction of Ce(IV) to Ce(III) with the concomitant release of $H^+$ ion and oxygen molecule. Subsequently, another molecule of hydrogen peroxide interacts with a vacant oxygen site (with high $Ce^{3+}$ ions), which results in the formation of Ce(IV) with the concomitant release of a water molecule. The reaction mechanism (see below) is analogous to the iron present in natural CAT[68].

$$H_2O_2 + Fe(+3)\text{-}E \rightarrow H_2O + O{=}Fe(+4)\text{-}E^+$$
$$H_2O_2 + O{=}Fe(+4)\text{-}E^+ \rightarrow H_2O + Fe(+3)\text{-}E + O_2$$

**2.1.3 Hydroxyl radical scavenging ability:** Hydroxyl radicals are highly reactive species generated during oxygen metabolism. However, there is no specific antioxidant enzyme in biological systems to scavenge these radicals. Some of the common enzymes such as SOD, CAT, glutathione peroxidase, melatonin and vitamin E are reported to balance the level of hydroxyl radicals in cells indirectly, and only a few reports have shown that nanoceria can directly scavenge hydroxyl radicals[69]. Using methyl violet as the chromogenic substrate, it was proposed that upon the reaction of hydroxyl radicals, Ce(III) is oxidized to Ce(IV) [69a]. Subsequently, regeneration of Ce(III) occurs due to several surface reactions involving hydrogen ions (from solution) and Ce(IV) (from nanoceria surface). The crystallite size and number of Ce(III) sites at the nanoceria surface are directly proportional to the hydroxyl radical scavenging property.

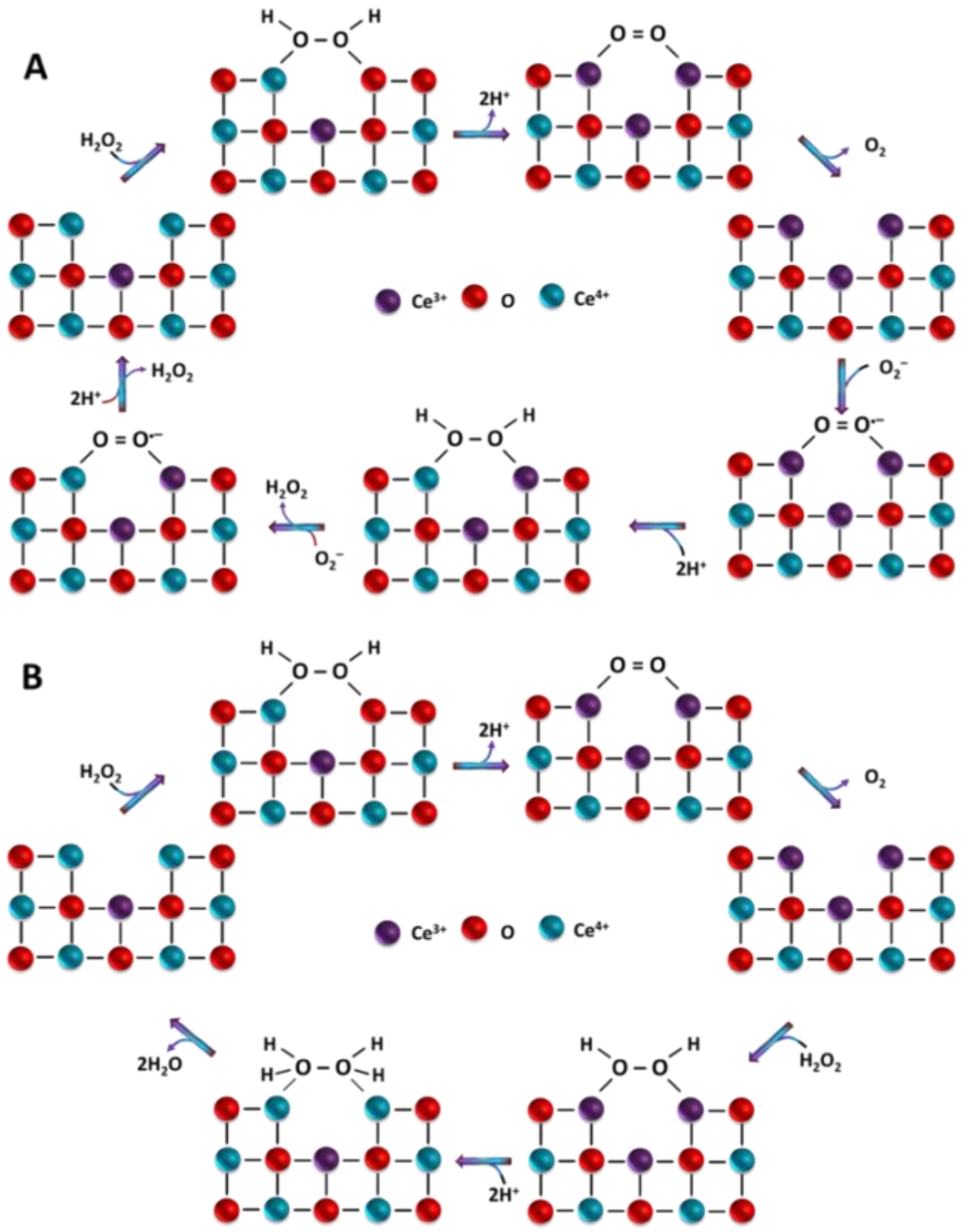

*__Figure 3:__ A schematic representation showing the possible reaction mechanism and alteration in the oxidation states of "Ce" atoms from nanoceria during the complete dismutation of superoxide radicals (A) and hydrogen peroxide (B). (Redrawn from – ref 57)*

**2.1.4 Reactive Nitrogen Species Scavenging Activity**: Nitric oxide is another reactive radical that combines readily with superoxide radicals to produce peroxynitrite (ONOO-). Nitrosonium cation (NO+), nitroxyl anion (HNO-), S-nitrosothiols (RSNOs), and dinitrosyl iron complexes are some other common and highly reactive nitrogen species in biological systems. Although less explored, nanoceria has also been shown to scavenge the RNS species effectively. It was reported that nanoceria with a low Ce(III)/Ce(IV)oxidation state ratio exhibited strong scavenging of nitric oxide radicals[70]. S-nitroso-N-acetylpenicillamine (SNAP) was used to generate nitrogen free radicals, which lead to the conversion (oxidation) of $Fe^{2+}$ ions (from haemoglobin) in $Fe^{3+}$.Nanoceria with low Ce(III)/Ce(IV) also significantly inhibited the oxidation process in a concentration-dependent manner. Subsequently, nanoceria was also found to accelerate the decay of the formation of peroxinitrites[71]. Interestingly, both Ce(III) and Ce(IV) nanoceria showed almost similar decay kinetics when followed using the oxidation of 3'-(paraaminophenyl) fluorescein (APF). Upon binding with peroxynitrite, APF shows a strong fluorescence (Ex/Em = 490/515 nm). However, the emission intensity is significantly inhibited in the presence of nanoceria similar to glutathione (a well-known scavenger of peroxynitrite).

**2.2 Pro-oxidant catalysis by nanoceria**: The term "pro-oxidant" refers to the reactions that result in the production of free radicals. There have been reports on the pro-oxidant nature of nanoceria, similar to the natural peroxidase and oxidase enzymes. A scheme of pro-oxidant and anti-oxidant reactions shown by nanoceria stemming from the reaction of $O_2{\bullet}^-$ are shown in **Figure 4**.

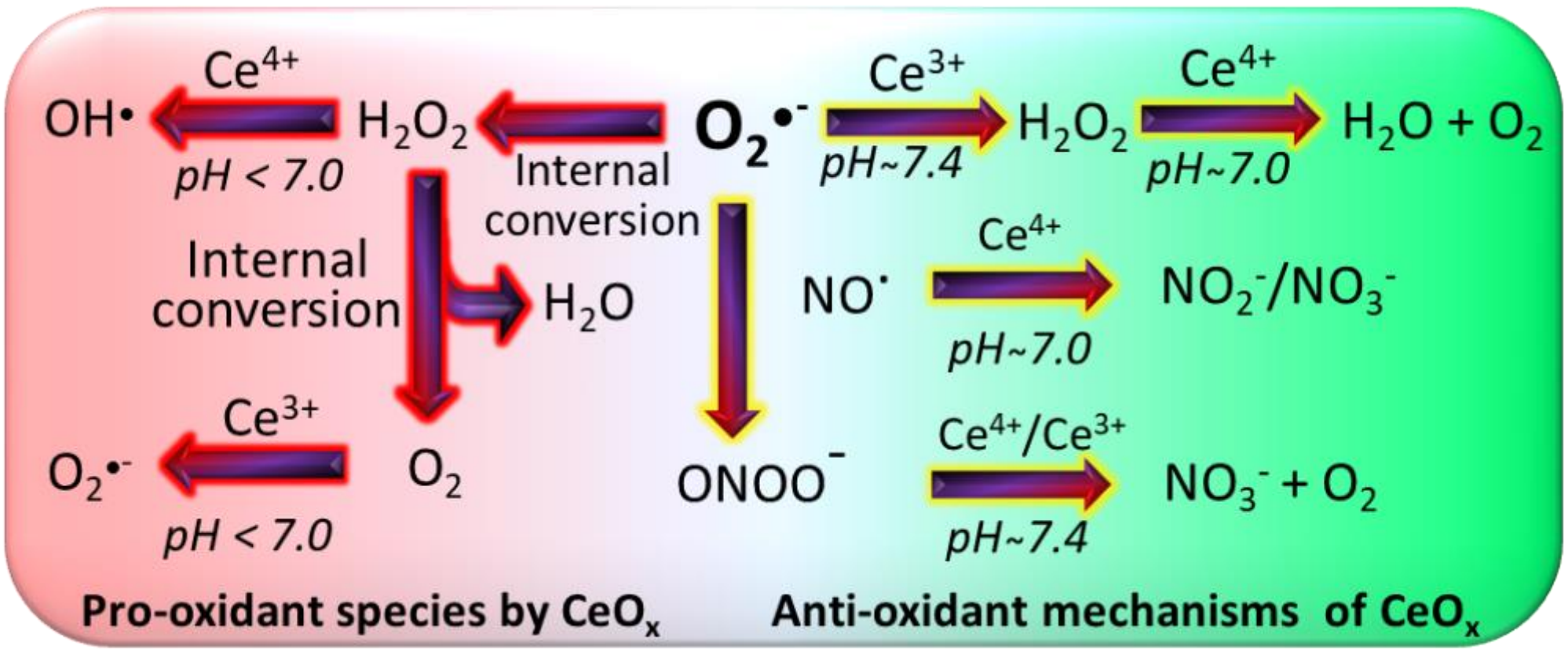

***Figure 4**: Schematic diagram showing the reaction of superoxide radicals with NO or $2H^+$ to produce ONOO− anions and hydrogen peroxide, respectively. The SOD mimetic enzyme or nanoceria (Ce(III)) can scavenge superoxide radicals into hydrogen peroxide that would subsequently be decomposed into water and molecular oxygen by CAT enzyme or nanoceria (Ce(IV)) at neutral pH. At neutral pH, nanoceria (Ce(IV)) can scavenge NO• radicals into nitrate/nitrite. However, at acidic pH, the hydrogen peroxide (generated via internal enzymatic pathways) can be degraded into hydroxyl radicals in the presence of peroxidase enzyme or nanoceria (Ce(IV)). Similarly, nanoceria (Ce(III)) can produce superoxide radicals at acidic pH from the oxygen generated generated through $H_2O_2$ degradation by internal enzymes.*

**2.2.1 Oxidase-like activity**: Biological oxidases catalyze the oxidation-reduction reaction where oxygen acts as an electron acceptor. Subsequently, the oxygen molecules are reduced, and the donation of a hydrogen atom leads to hydrogen peroxide and water production. Oxidases play key roles in the homeostasis of the body by mediating the use of oxygen for energy generation by controlling the flow of electrons in the electron transport chain. Perez and co-workers first reported the oxidase-like activity of nanoceria by showing the ability to oxidize the organic substrates such as TMB ((3',3',5',5'-tetramethylbenzidine) and ABTS (2,2-azinobis (3-ethylbenzothizoline-6-sulfonic acid)), without the need for hydrogen peroxide[30, 60b, 72]. Folate conjugated nanoceria was used to develop an immunoassay for the detection of folate expressing cancer cells. Subsequently, it was reported that the oxidase-like activity of nanoceria can be improved (100 fold) by the use of fluoride[29a]. Being a hard Lewis base, fluoride shows a strong affinity for cerium (hard Lewis acid), strongly interacts with nanoceria and favourably modulate the oxidase-like activity. Mechanistically, it was hypothesized later that the oxygen is adsorbed on the surface of nanoceria, leading to the formation of $O_2{\bullet}^-$ (in acidic condition). Next, Ce(IV) is reduced to Ce(III) with concomitant oxidation of substrate followed by oxidation of Ce(III) by $O_2{\bullet}^-$ into Ce(IV). The summary of the overall reaction can follow the below given sequence of reactions[60a]; however, evidence of the involvement of Ce(III) and Ce(IV) by varying the concentration of surface Ce(III)/Ce(IV) ratio is missing in these studies.

$$Ce^{3+} + O_2 \rightarrow Ce^{4+} + O_2^{\bullet -}$$
$$O_2^{\bullet -} + \text{Substrate(red.)} \rightarrow H_2O + \text{Product(ox.)}$$
$$CeO_2 + \text{Substrate(red.)} \rightarrow Ce_2O_3 + \text{Product(ox.)}$$
$$Ce_2O_3 + O_2^{\bullet -} + 2H^+ \rightarrow CeO_2 + H_2O$$

**2.2.2 Peroxidase-like activity**: Biological peroxidases are a large group of ubiquitous enzymes that catalyse the break down of peroxides. These are divided into heme-based and non-heme based proteins that are further divided into multiple types based on the type of active sites and residues. These enzymes classify under the oxido-reductases group catalyzing the oxidation of the substrates by decomposition of hydrogen peroxide. Due to the redox-active behavior, nanoceria has also been reported to exhibit peroxidase-like activity, which was utilized for hydrogen peroxide and glucose detection within the range from $6.0 \times 10^{-7}$ to $1.5 \times 10^{-6}$ mol $L^{-1}$ and $6.6 \times 10^{-6}$ to $1.3 \times 10^{-4}$ mol $L^{-1}$, with the detection limit down to $5.0 \times 10^{-7}$ mol $L^{-1}$ $H_2O_2$ and $3.0 \times 10^{-6}$ mol $L^{-1}$ glucose, respectively[73]. Doping of transition metals such as Mn, Fe, Co, Ni and Cu has been shown to modulate the peroxidase mimetic activity of nanoceria with Mn doping giving the highest acitivity[74]. Analogous to Fenton/Haber-Weiss reactions, nanoceria has also been proposed to follow a similar reaction mechanism mediated by hydroxyl radicals to exhibit peroxidase mimetic activity[30]. Interestingly, Heckert *et al*. have investigated the Fenton-like reaction shown by cerium ions in the presence of hydrogen peroxide[75]. The results revealed that the use of cerium chloride in the presence of hydrogen peroxide leads to the generation of hydroxyl radicals that were assessed by causing relaxation in supercoiled plasmid DNA. Further, the oxidation of peroxidase substrate (AzBTS) and the use of spin traps revealed that both hydroxyl radical and superoxide anions are generated in the reaction containing cerium element and hydrogen peroxide.

As described in this section, nanoceria can show multiple enzyme like activities that have specific advantages as compared to the naturally existing enzymes. The high surface area and small size of nanoceria presents a large number of active sites on the surface of a single nanoparticle for scavenging of ROS and RNS as compared to natural enzymes that usually have one or two metal centers/active sites for performing their action. Nanoceria shows much higher tolerance to subtle changes in the temperature and pH as compared to natural enzymes. Most enzymes are localised or cannot be present at different locations on demand for ROS scavenging while the high cellular permeability nanoceria ensures that it can be present in suitable concentration at desired locations. Lastly, nanoceria can also show multiple enzyme like activity that can be tuned by the addition of specific ligands or due to changes in the pH as compared to natural enzymes that are very specific for a particular application. While there are specific advantages, nanoceria also has certain drawbacks as compared to natural enzymes. One of the major drawback of nanoceria (and most of the nanomaterials based enzymes) is the lack of specificity towars a particular reaction. Thus, nanoceria is not active towards a specific ROS or RNS species and demonstrates a broad spectrum of activity. Self and co-workers have demonstrated that nanoceria can scavenge superoxide radicals with comparable kinetics to that of the SOD enzyme however, the peroxide scavenging activity of nanoceria is slower as compared to natural enzymes such as the CAT. Even though nanoceria can depict oxidation state dependent activity however, this can also be deleterious, such as, the oxidase-like activity shown by nanoceria with lower Ce(III)/Ce(IV) ratio at acidic pH can demonstrate cytotoxic behaviour in some cells that does not happen in naturally existing enzymes. Lastly, the biological activity of nanoceria can vary based upon the sysnthesis method due to changes in surface oxidation state, surface defects and geometry while natural enzymes have fixed structures and thus a defined structure-activity relationship.

On the basis of the above discussion, it can be concluded that nanoceria has specific advantages and disadvantages as compared to naturally existing enzymes. It also suggests that the redox-based cycling of Ce(III) and Ce(IV) has been well-described in the literature, but the role of capping agents on the redox behavior has not been well investigated except some of the very recent investigations that are described in the next section.

### 3.0 Impact of intentional and unintentional surface coatings on the catalytic activity of nanoceria

Over the past 20 years of the development of nanocarriers for medical use, reports have shown that it is crucial to add surface functionalities to the as-synthesized particles and assess their efficiency with respect to their colloidal stability and the formation of the protein/chemical corona. The sequence *synthesis – functionalization - stability* holds in principle for all types of newly developed nanostructures. For nanoceria, it is also necessary to verify that the redox properties are not altered by the coating. To date, there are a number of promising studies that show that polymer-based coatings can preserve both the catalytic and dispersibility properties of nanoceria in biological media.

### 3.1 Polymer chains

A broad range of techniques in chemistry are now being applied to the coating and stabilization of nanoceria. A widely used coating method is the *one-step synthesis and functionalization process*, in which low molecular weight ligands[76], or polymers[72, 76b, 77] are introduced with cerium salts during the nanoceria synthesis (**Table 2**). The organic moieties are added to control the nucleation and growth of the $CeO_2$ nanocrystals, and at the same time, to provide colloidal stability. Another efficient pathway is the *two-step post-synthesis process*, where the coating is bound to the nanoceria surfaces after $CeO_2$ synthesis (**Table 3**). This can be done alternatively using surface-initiated living polymerization, also called *grafting-from* technique [78] or the co-assembly *grafting-to* technique[29b, 79]. This later strategy includes the ligand or polymer adsorption, a technique similar to layer-by-layer deposition[29b, 79a, 79b, 79f, 79h, 79l] and the phase transfer of coating agents between organic and aqueous solvents[79c, 79d, 79g].

Pertaining to the *one-step synthesis and functionalization process*, Perez *et al.* synthesized 4 nm dextran coated $CeO_2$ particles that presented a color change from yellow to orange upon $H_2O_2$ addition, which was later shown to vanish over time[72]. This outcome was interpreted in terms of autocatalytic behavior induced by the reversibly switch from Ce(III) to Ce(IV) oxidation states at the particle surface. It also showed that the dextran coating did not affect the antioxidant properties of nanoceria. Similar results on the $H_2O_2$-mediated redox cycling, together with evidence of SOD mimetic activity, were later demonstrated by Karakoti *et al.* on a series of PEGylated nanoceria (**Figure 5**)[77a]. Nanoceria coated with dextran have recently found application in nanomedicine as contrast agents in computed tomography for the detection and protection of inflammation sites *in vivo*, a result that had not been highlighted until now[77e]. Numerous other studies have relied on the *one-step synthesis and functionalization process* for nanoceria formulations that were mainly assessed with respect to their cytotoxicity and protection against cellular ROS[77c, 77d, 77f]. One of the issues related to the above synthesis method is that the effects of the coating on the antioxidant properties cannot be directly inferred. With the *two-step post-synthesis process*, nanoceria and coatings are synthesized separately and later assembled using appropriate protocols. This technique mainly takes advantage of the extended library of functional polymers developed via polymer/coordination chemistry.

**Table 2: Synthesis methods, associated redox activity, and colloidal stability of nanoceria**

| Synthesis method | Size by TEM or XRD (nm) | Type of coating material | Protein corona or colloidal stability | Redox properties | Ref |
|---|---|---|---|---|---|
| **Alkaline-based precipitation** | 4 | Dextran | Not studied | - Autocatalytic activity<br>- ROS generation at acidic pH | [72] |
| **Oxidation of cerium sols using $H_2O_2$** | 3 – 5 | PEG 600 Da | Not studied | - SOD activity<br>- $H_2O_2$ redox cycling<br>- Sod-activity increases with PEG | [77a] |
| **Microwave–hydrothermal treatment** | 2-3 | Citric acid<br>Poly(acrylic acid) 8 kDa | - PAA protects from aggregation<br>- Citric acid, CNP coarsening with temperature | Not studied | [76b] |
| **Alkaline-based precipitation** | 3 - 4 | Dextran 10 kDa | Not studied | Dextran coating improves cell viability | [77d] |
| **Alkaline precipitation** | 3-4 | Dextran 10 kDa | - Formation of protein corona with FBS<br>- CNP size depends on pH | - CNP@Dextran show pH dependent ROS generation (High cytotoxicity for cancer cells at pH 6, non-toxic for healthy cells at pH 7) | [77c] |
| **Alkaline precipitation** | 4 | Ethylene glycol | CNP@EG aggregates in cell media | - CNP@EG protects against $H_2O_2$ induced ROS | [76a] |
| **Wet chemical synthesis** | 2.5 | Citric acid<br>EDTA<br>Citric acid/EDTA | Not studied | - SOD and CAT activity<br>- CNP@CA/EDTA has potent antioxidant effects | [77b] |
| **Alkaline precipitation** | 4.8 | Dextran 10 kDa | Not studied | - Antioxidant activity in cell culture | [77e] |

coated by various materials

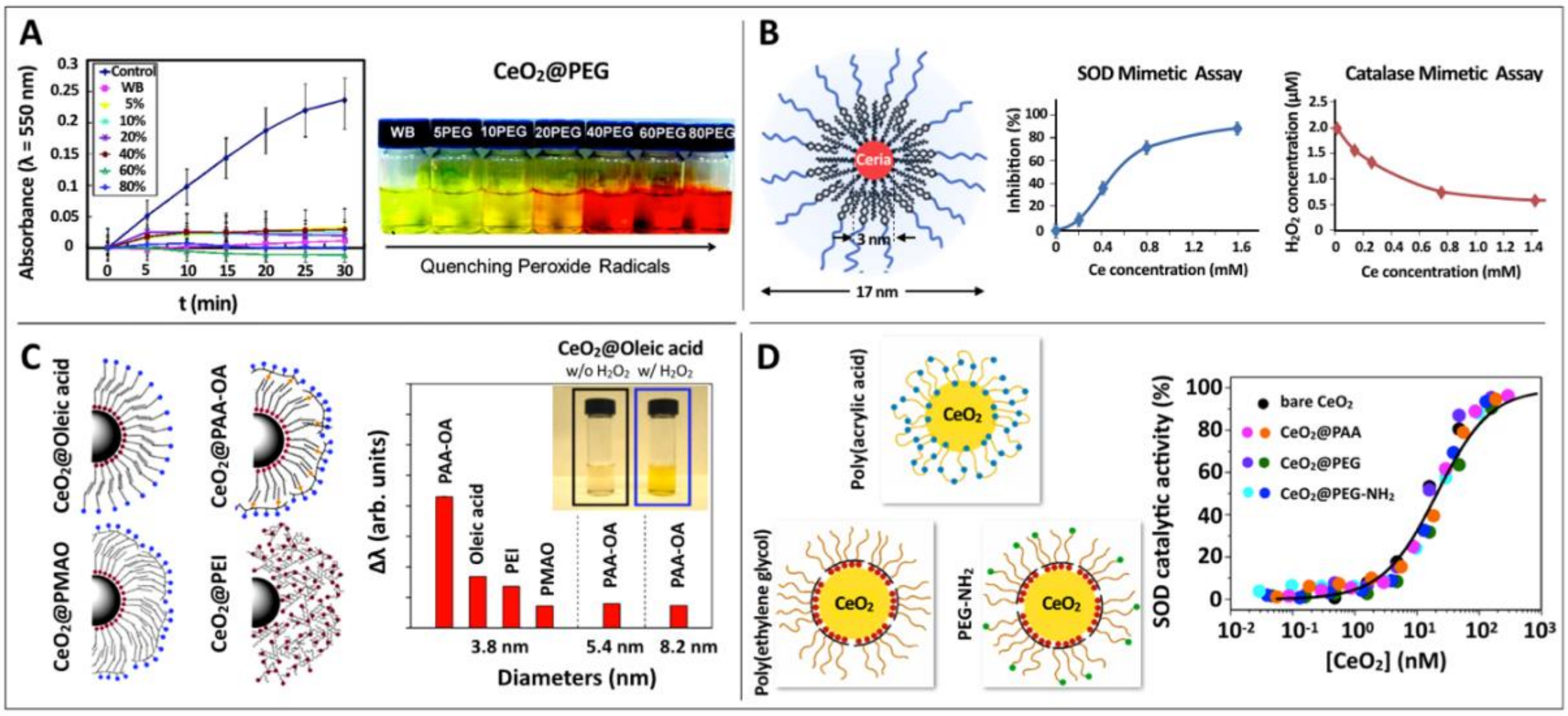

***Figure 5: A)*** *Dispersion of 3-5 nm $CeO_2$ NPs coated with PEG 600 Da obtained from one-step synthesis and functionalization technique. The superoxide dismutase (SOD) mimicking activity is measured through the time dependence of the absorbance after $H_2O_2$ addition (right panel) [Reproduced with permission from Karakoti et al.[77a]].* ***B)*** *Nanoceria geometry after hydrophilic encapsulation with lipid-PEG used as transfer agents between the organic and aqueous solvents (left panel). Concentration dependences of the SOD and catalase mimetic activity for the lipid-PEG CNPs [Reproduced with permission from Kim et al.[79c]].* ***C)*** *$CeO_2$@Oleic acid, $CeO_2$@PAA-OA, $CeO_2$@PMAO, and $CeO_2$@PEI coated nanoceria obtained from phase transfer between organic and aqueous solvents. The red shift of the absorbance $\Delta\lambda$ measured at the optical density 0.30 after injection of $H_2O_2$ is a marker of the CNP catalytic activity [Reproduced with permission from Lee et al.[79d]].* ***D)*** *Representation of 7 nm CNPs coated with poly(acrylic acid) ($CeO_2$@PAA), with multiphosphonic acid PEG copolymers ($CeO_2$@PEG) and with the same copolymers where the PEGs are terminated by an amine group ($CeO_2$@PEG-$NH_2$). Percentage of dismutated superoxide radicals obtained for bare $CeO_2$, $CeO_2$@PAA, $CeO_2$@PEG and $CeO_2$@PEG-$NH_2$. As a function of the $CeO_2$ molar concentration. The continuous curve results from a Langmuir-type adsorption isotherm model [Reproduced with permission from Baldim et al.[79a, 79b]].*

With uniform 3 nm nanoceria in chloroform, Kim *et al.* used PEGylated lipids as transfer agents between the organic and aqueous solvents, resulting in core-shell particles with good stability in buffers and blood plasma. The particles were shown to retain their SOD, CAT and autocatalytic activities, to scavenge ROS in vitro and protect brain damage following ischemic stroke[79c]. Later, the same group developed zirconium doped nanoceria with improved ROS scavenging performance to regulate inflammatory cells at lower doses[79g]. With the same coating technique, Lee *et al.* examined a series of organic coatings (including oleic acid, poly(acrylic acid), poly(ethylene imine) of thicknesses between 2 and 8 nm, and found that the presence of a surface coating did not preclude oxidation reactions, and that the most reactive nanoceria was that with the thinnest surface coating, here obtained with oleic acid[79d].

Table 3: Post-synthesis coating using low molecular weight ligands or polymers

| Synthesis | Size (nm) Morphology | Coating material | Solvents used | Protein corona or colloidal stability | Redox properties | Ref |
|---|---|---|---|---|---|---|
| **Oxidation of cerium sols using $H_2O_2$** | 11 Spherical | Phosphate, sulfate, carbonate | Electrolytes, PBS, DMEM | - Phosphate induced CNP aggregation | - Coated CNPs display SOD and CAT activity - Phosphates decrease SOD and increase CAT | [48d] |
| **Modified reversed micelle method** | 3 Spherical | PEGylated lipids | DI-water PBS DMEM | - Good stability in culture medium and blood plasma | - Coated CNPs display SOD and CAT activity - Autocatalytic activity - *In vitro* antioxidant activity | [79c] |
| **Thermal decomposition of cerium salt** | 3.8, 5.4 and 8.2 Spherical | Oleic acid PAA-Oleic acid PMAO and PEI | DI-water PBS | Not studied | - Coated CNPs display antioxidant activity - Activity increased for thin coating (oleic acid) | [79d] |
| **Sigma-Aldrich (CA289744)** | 5–8 Spherical | Citrate, Phosphate, DNA, PSS, PVS, $PAA_{8K}$, $PAA_{15K}$, $PEG_{2K}$, $PEG_{20K}$ | PBS | - DNA binds CNPs - CNPs aggregate in buffers (CA, phosphate, MES, electrolytes) | - DNA inhibits oxidase activity - Catalytic activity: increases with PAA and PEG, decrease with PSS and PVS | [29b] |
| **Thermo-hydrolysis of cerium salt** | 6.8 Agglomerates of 2 nm crystallites | Citrate, $PAA_{2K}$ | DI-water PBS DMEM | - CNP@CA aggregate in PBS and DMEM - CNP@PAA: high stability (> 1 month) in culture media | *In vitro* antioxidant activity | [79f] |
| **Plasma** | 15 Spherical | Polymers PSPM and PMETAC | Not specified | - Coatings improve colloidal stability - Evidence of protein corona in cell media | Antioxidant activity slightly lower with coating | [78] |
| **Flame spray pyrolysis** | 7 Faceted | Hyaluronan (170 kDa) *grafting from* approach | Not specified | Not studied | nanoceria@HA show intracellular antioxidant activity | [79e] |
| **Modified reversed micelle method** | $Ce_{0.7}Zr_{0.3}O_2$ - 2.2<br>$CeO_2$ - 3.6 | PEGylated lipids Coated NP 11-16 nm in DLS | DI-water<br>PBS<br>DMEM | - Colloidal stability in cell media and blood plasma<br>- No protein corona | - SOD, CAT, HORAC activities<br>- *In vitro* antioxidant activity<br>- $Ce_{0.7}Zr_{0.3}O_2$ is 10 times active than nanoceria | [79g] |
| **Hydrothermal reaction** | 5-20 Cubes<br><br>50 -100 Rods | $PEG_{5K}$ *grafting from* approach | Not specified | - $PEG_{5K}$ mitigates protein corona<br>- $CNP@PEG_{5K}$ aggregate in culture media | - Coated nanoceria display CAT activity | [79h] |
| **Wet chemical and thermal hydrolysis method** | 8-12 Spherical | Phosphine-based ligands<br>TEP and TTMPP | Not specified | Not studied | - Coated nanoceria display SOD and CAT activity<br>- $CeO_{2-x}$ more efficient than $CeO_2$ | [79j] |
| **Thermo-hydrolysis of cerium salt** | 6.8 Agglomerates of 2 nm crystallites | Mono- and multiphosphonic acid PEG copolymers (2 and 5 KDa) | DI-water<br>PBS<br>DMEM | Multiphosphonic PEG coating prevent aggregation and protein corona | Not studied | [79b] |
| **Thermo-hydrolysis of cerium salt** | 6.8 Agglomerates of 2 nm crystallites | Multiphosphonic acid PEG copolymers | DI-water<br>Tris and acetate buffers | Coated CNPs are stable in tris and acetate buffers | Coating lowers SOD-like, preserve CAT-like and increase peroxidase activities | [79a] |

The tendency of synthetic polymeric coatings to retain nanoceria antioxidant properties was later confirmed using co-assembly[29b, 79a, 79b, 79e, 79f, 79h] or the in-situ polymerization[78] methods, both in acellular and *in vitro* conditions[48d, 72, 76a, 77c, 77d, 79d-f, 80]. The trend was found not only for spherical nanoceria, but also for rods and cubes, as exemplified by the work of Xue *et al.*[79h]. However, most studies to date have not investigated the relationship between the structure of the polymer layer (which can be described in terms of number density of attached chains or spatial monomer distribution) and the changes in redox properties. This knowledge is expected to be key for the control of the final nanoceria properties. In this context, Baldim *et al.* investigated a series of phosphonic acid PEG copolymers that were grafted to 7.8 nm nanoceria *via* condensation and/or coordination with cerium ions [79a, 79b]. Keeping the nanoceria core identical and varying the coating layer, these authors could show that compared to bare nanoceria, the robust PEG layer did not affect the SOD-like activity and impaired the CAT-like activity by 30%. Surprisingly, this coating improved the peroxidase-like catalytic activity of nanoceria by a factor of 2-3, a property that is likely to arise from the synergistic action between the polymers and the particles. It was concluded that PEGylated brushes of density 0.2 – 0.5 $nm^{-2}$ and height 10 nm preserve the accessibility of free radicals to the reactive sites, while ensuring excellent stability of particles in biological fluids[79a].

### 3.2 Low molecular weight ligands

While the data currently available on polymer coatings show an overall and clear trend, the results on the low molecular weight ligands adsorbed on the nanoceria surface are more contrasted. Singh *et al.* have observed a complete loss of SOD activity of uncoated nanoceria after incubation with phosphate buffer at physiological concentration. The SOD and CAT catalytic behavior shift was explained by the strong binding of phosphate anions to cerium [48d]. Given the abundance of phosphate in biological systems, such a result could hamper the translation of nanoceria as therapeutic agents. The strong affinity of phosphate groups with cerium oxide surfaces was also disclosed using single and double strand DNA oligonucleotides [29b]. When adsorbed on bare nanoceria, the enzyme-like oxidase activity was inhibited, a result that was explained again by the tight binding of the DNA phosphate backbone to cerium through Lewis acid-base interaction. Pautler *et al.* also reported that the oxidase enzyme-mimetic activity was retained with phosphate, citrate and acetate moieties adsorbed at their surface[29b]. Recently, Patel *et al.* found a reverse effect when the nanoceria were coated with low molecular weight phosphine ligands[79j]. In one example studied, that of triethyl phosphite, the presence of ligands at the nanoceria surfaces increased noticeably the SOD response with respect to that of the uncoated particles. Taken together, these results suggest that nanoceria reactivity is high with low molecular weight ligands, and that depending on the electronic properties and of the density of the adsorbed molecules, the antioxidant properties can be profoundly altered, either upward or downward. The conformation of highly solvated polymer chains, in contrast, allows a less dense coverage of the nanoceria surfaces, promoting the accessibility of ROS-like species towards the active redox sites. There is a general lack of understanding in the role of low molecular weight ligands on the properties of nanoceria. This is an upcoming area of research that can have far reaching implications on nanoceria research including modulating its specificity and activity. Specifically, this area of research can also benefit from the use of molecularly imprinted polymers (MIPs) to improve the substrate specificity of nanoceria based nanozymes. MIPs are designed to mimic the structure specificity of antibody-antigen systems by following a template assisted design approach in which the target molecule or substrate can act as the template[81]. Such advanced techniques could pave the way for reforming the field of nanozymes and bringing them closer to the function of natural enzymes.

### 4.0 Applications of nanoceria in various disease models

Nanoceria has been explored in a range of disease models where oxidative stress is a major contributor to the pathogenesis of the condition. Nanoceria is usually delivered intravenously or intraperitoneally for therapeutic application, in contrast to the intratracheal administration explored in accidental environmental release models. Intratracheal administration of nanoceria in rodent models at high doses up to 400 mg/kg increased ROS as well as pro-inflammatory cytokine production, DNA damage, granuloma formation and fibrosis[82]. These findings contrast with the topical, intravenous, intraperitoneal or intravitreal administration at doses up to 0.5 mg/kg in multiple disease models where nanoceria reduces oxidative stress and pro-inflammatory cytokine release and promotes tissue repair. Thus, the dose and route of administration are likely crucial determinants of the biological effect of nanoceria as detailed in the examples below. The use of nanoceria as a drug release model system also necessitates the importance of understanding the biological effects of nanoceria in these disease models[80c, 83].

**4.1 Systemic inflammatory response syndrome** is the primary cause of death in septic patients and is characterized by elevated ROS and pro-inflammatory cytokines, such as tumour necrosis factor (TNF)-α and interleukin (IL)-6. In humans, sepsis is most frequently caused by gram-positive organisms and fungi while animal models of sepsis largely utilize lipopolysaccharide (LPS), a component of the cell wall of gram-negative bacteria[84]. In addition, these models lack the infectious component that mounts the immune reaction characterized by the human condition[84]. Furthermore, rats are more resistant to LPS challenge than humans requiring high doses that induce toxic effects with an acute and transient increase in inflammatory cytokines, in contrast to the gradual and prolonged increase seen in humans[85]. Thus, the model more closely resembles endotoxic shock than sepsis, however, both conditions involve oxidative stress making this model applicable to studies of nanoceria[86]. Nanoceria treatment of LPS-induced sepsis in rats reduced serum levels of pro-inflammatory cytokines including TNF-α, IL-1β and IL-1α[87]. Liver damage was evident in this model via both histological analysis and serum markers of liver function with elevated serum levels of pro-inflammatory cytokines largely produced by Kuppfer cells in the liver, while administration of nanoceria reversed these changes[87]. The cecal inoculum rat model of systemic inflammatory response syndrome causes intra-abdominal infection and acute kidney injury. The administration of nanoceria in this model conferred protection against acute kidney injury as measured by serum kidney function levels and histology (**Figure 6A**)[88]. This model has also been used to demonstrate that nanoceria reduced proteolysis, pro-inflammatory signalling and improved diaphragm function, which is associated with a better prognosis in humans[89]. Similar findings were reported for 6-aminohexanoic acid surface modified nanoceria in the cecal ligation model of sepsis[90].

**4.2 Critical limb ischemia** involves the loss of blood flow, impaired wound healing and tissue necrosis. It is characterized by an overproduction of mitochondrial ROS primarily due to hypoxia causing proinflammatory cytokine release. Intramuscular administration of nanoceria in a hindlimb ischemia model in immunodeficient nude mice supported revascularization via upregulation of angiogenic growth factors including hypoxia induced factor (HIF)-1α and apurinic/apyrimidinic endonuclease 1 (APE1), an enzyme which repairs oxidative base damage[91]. Similarly in ischaemic stroke, disruption to cerebral blood flow and oxygen provides a hypoxic environment where oxidative stress results in vasodilation, breakdown of the blood brain barrier and neuronal cell death[92]. Administration of nanoceria in the subarachnoid hemorrhage rat model of stroke reduced neuronal death, macrophage infiltration and brain oedema and improved survival,[93] suggesting the ability of nanoceria to provide disease-modifying therapies.

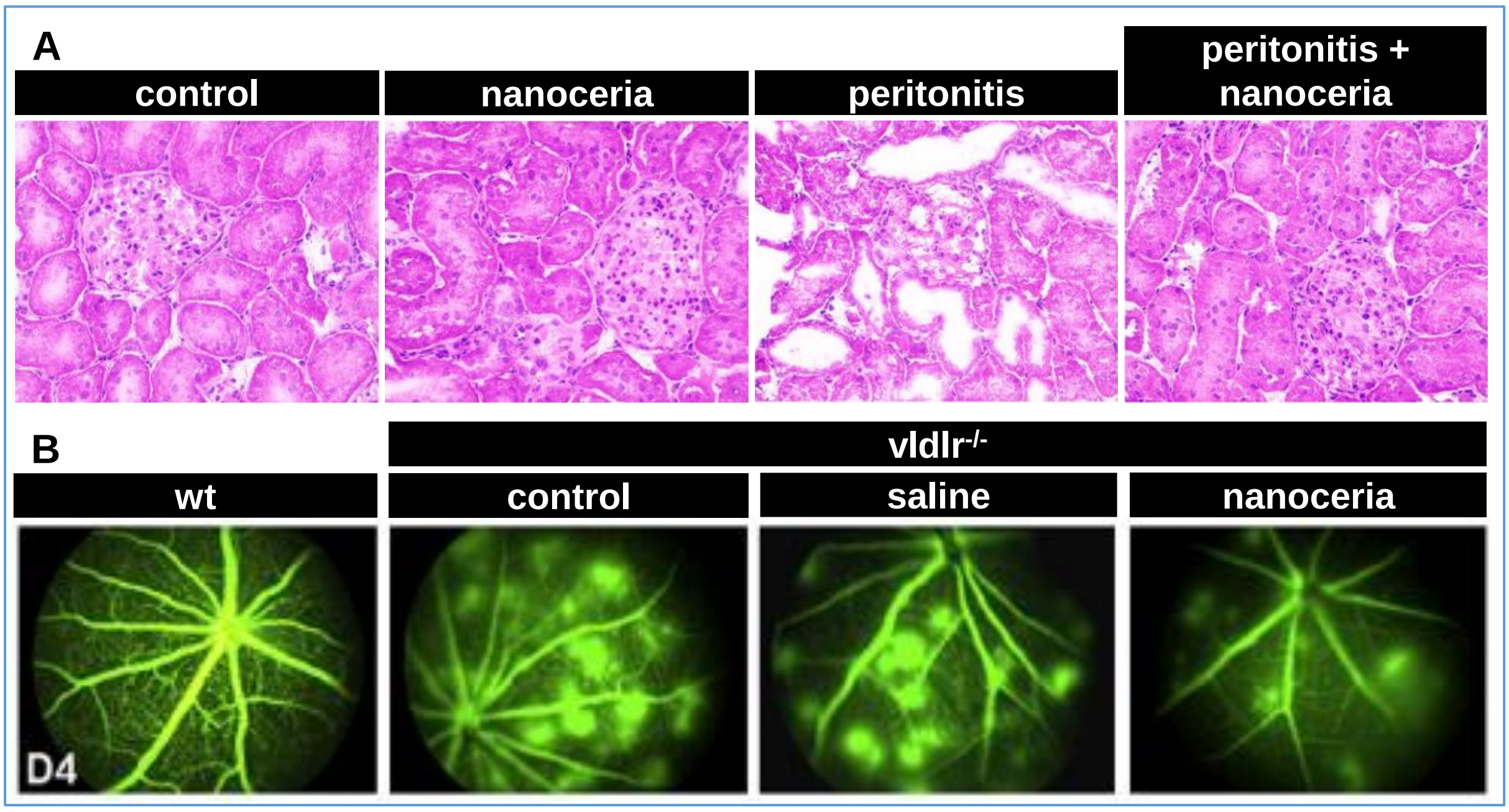

***Figure 6**. **(A) Nanoceria attenuate peritonitis-induced renal damage in a rat model of acute kidney injury.** Hematoxylin and eosin staining of kidney sections 18 h after treatment with nanoceria with or without polymicrobial insult (peritonitis). Peritonitis was characterized by renal tubular dilatation, loss of brush border and damage to the glomerular capillary network in this model while intraperitoneal administration of 0.5 mg/kg nanoceria following polymicrobial insult attenuated these changes. [Reproduced with permission from [88]]. **(B) Nanoceria regress neovascularization and prevent leakage in very low-density lipoprotein receptor knockout (vldlr-/-) mouse model of human retinal angiomatous proliferation, a form of wet age-related macular degeneration.** Fluorescent angiography revealed numerous fundus patches of neovascularization and blood vessel leakage in the vldlr-/- mice compared to wild-type (wt). Intravitreal 1 mM nanoceria injection reduced neovascularization and leakage in the vldlr-/- mice compared to control or those treated with saline. Reproduced with permission from[94].*

**4.3 Diabetes mellitus** is a chronic condition associated with increased blood glucose levels due to impaired insulin secretion or function. Hyperglycaemia leads to advanced glycation end products that generate free radicals and oxidative stress, tissue damage and delayed wound healing[95]. The streptozotocin-induced rat model of diabetes is the most commonly used model to recapitulate the human condition, although its natural level of pancreatic regeneration is not seen in humans[96]. Wound healing is often impaired in diabetes and causative for most peripheral limb amputations. Thus, therapies that promote cutaneous wound healing have been widely investigated. Nanoceria offers the potential to modulate oxidative stress and when delivered in an electrospun poly(3-hydroxbutyrate-*co*-3-hydroxyvalerate) membrane to a full-thickness excisional wound in the streptozotocin rat model, enhanced cell infiltration and granulation tissue formation, suggesting a beneficial effect[97]. Similar findings have been reported for topical delivery of nanoceria to full thickness wounds in healthy rodent models with enhanced wound healing, and leukocyte infiltration observed[98]. These findings suggest that nanoceria is a promising antioxidant to further explore for wound healing applications. While this model enables assessment of acute wound healing, extrapolation to human cutaneous wound healing is confounded by rapid wound contraction and differences in the immune system[99].

The streptozotocin-induced diabetic rat model has also been used to explore **diabetic neuropathy** which is mediated by mitochondrial oxidative stress[100]. In this model, neuron damage was decreased with intraperitoneal nanoceria treatment[101]. Similar findings were observed in a streptozotocin-induced diabetic mouse model, which displayed reduced glucose and pro-inflammatory cytokine expression[102].

Oxidative stress is tightly linked to mitochondrial dysfunction, which is a major cause of **neuronal dysfunction and degeneration** and underlies a variety of vision impairments. Nanoceria can reside within the eye for extended periods when delivered intravitreally in albino rats[103], suggesting that they may be of therapeutic value. Retinitis pigmentosa is a group of hereditary retinopathies characterized by photoreceptor cell death and retinal degeneration. While this is a genetic condition, there is evidence of oxidative stress and altered autophagy[104]. P23H-1 rats are a model for retinitis pigmentosa as they carry a point mutation in the rhodopsin gene. Intravitreal administration of nanoceria in this model provided neuroprotection by slowing photoreceptor cell death[105]. The tubby mouse exhibits obesity at maturity, which is accompanied by retinal degeneration and is caused by a mutation of the *tub* gene, which encodes for a transcription factor involved in protein trafficking. Loss of this transcription factor alters the location of proteins involved in photoreceptor function involving oxidative stress. Administration of nanoceria in this model enhanced photoreceptor survival with long-term activity[106].

**4.4 Age-related macular degeneration** (AMD) also involves photoreceptor death and microglia activation. The very low-density lipoprotein receptor knockout (*vldlr -/-)* mouse is a model for human retinal angiomatous proliferation, a form of wet AMD. It is characterized by progressive neovascularisation of the retina, disruption of the retinal pigment epithelium as well as photoreceptor degeneration and fibrosis. Oxidative stress and inflammation induce the expression of vascular endothelial growth factor (VEGF) to promote neovascularization in this condition. Intraocular administration of nanoceria in the *vldlr*$^{-/-}$ model reduced neovascularization and VEGF expression and prevented photoreceptor death[94] (**Figure 6B**). In addition, intraocular administration of nanoceria in older mice enabled regression of vascular lesions, indicating that pathologic vessels require a continual supply of ROS-mediated factors for their survival[107]. In addition, this study indicated that nanoceria might be an effective disease-modifying therapeutic. Light damage to the dorsal region of the retina of albino rats causes oxidative stress-induced photoreceptor cell death as in the human condition. Intravitreal administration of nanoceria in this model localised to the retinal ganglion cell layer and persisted for 3 weeks, while intravenous administration of nanoceria only persisted in this region for 24 h. In addition, nanoceria reduced TNF-α and FGF2 levels with the preservation of visual function[108]. A disadvantage of rodent models is their lack of a macula, limiting their ability to replicate the aetiology of the human AMD. However, these models have shed light on the mechanisms of oxidative damage in this condition and thus are likely to provide valuable insight into the mechanisms by which nanoceria modulates these processes in disease[109].

**4.5 Radiation therapy** is commonly used in cancer treatment with radiosensitizers used to increase the susceptibility of cancer cells to this treatment whilst minimizing collateral damage to surrounding healthy tissue. Nanoceria conferred protection from X-ray irradiation to mice with 60 % survival compared to when treated with irradiation alone[110]. Nanoceria also conferred radioprotection for healthy lung tissue in athymic nude mice and rats when delivered via intraperitoneal injection, while pneumonitis and fibrosis were observed in untreated animals [111]. Similar effects were observed for the gastrointestinal epithelium in this mouse model[112]. Additionally, nanoceria provided radioprotection to healthy tissue in the athymic nude mouse model, but selectively sensitised human pancreatic cancer cells via induction of ROS and tumour cell apoptosis[113]. Similarly, nanoceria treatment of a human ovarian cancer xenograft in nude mice reduced tumor burden and potentiated cisplatin toxicity[114]. Administration of diethylnitrosamine in mice is a hepatocarcinogenesis injury model which is established to cause oxidative stress[115]. Nanoceria treatment reduced markers of oxidase activity, including myeloperoxidase and nitric oxide as well as SOD and CAT levels with reduced inflammatory cell infiltration compared to controls[116].

**4.6 Tissue regeneration** is the goal of many biomaterial-based therapeutic strategies with nanoceria able to promote key pathways in tissue regeneration such as angiogenesis. The liver has an innate ability to repair with the partial hepatectomy model, a popular choice to study due to its absence of complications associated with inflammation. Pre-treatment with nanoceria prior to partial hepatectomy promoted liver regeneration in this model in rats with more proliferative hepatocytes and reduced serum liver function measures [117]. Nanoceria also has positive effects on bone regeneration models. For example, PEG-coated nanoceria co-delivered with bone marrow stem cells in the subcutaneous ectopic osteogenesis model in nude mice accelerated new bone formation and endochondral ossification[118]. In addition, nanoceria modified bioglass scaffolds used in the rat cranial defect model enhanced bone regeneration[119].

This body of evidence suggests the ability of nanoceria to impart antioxidant properties in different disease models. While these disease models do not always completely replicate human conditions, they provide compelling evidence of the biological antioxidant capability of nanoceria, and without toxic side-effects, making this class of nanomaterial worthy of further pre-clinical development.

## 5.0 Burning issues in nanoceria research

### *5.1 Role of oxygen vacancies versus the oxidation state*

It is generally was believed that oxygen vacancies play a role in the antioxidant activity of nanoceria. Theoretically, it has been shown that oxygen adsorption on vacancies is an energetically favourable process. These oxygen molecules undergo reduction to form superoxide radicals by accepting an electron from the neighbouring Ce(III) cation[58]. Further conversion of superoxide to peroxide is fast and goes in a unidirectional manner, resulting in the oxidation of the nearest Ce(III) cation[58]. This reveals the importance of oxygen vacancies which indirectly assist the antioxidant mechanism of nanoceria. A classic study proposed by Celardo et al.[120] challenged this concept by depicting that the concentration of Ce(III) ions and not the oxygen vacancies is a key to the antioxidant activity of nanoceria. This was shown by doping the lattice of nanoceria with samarium ions (Sm(III)), replacing some of the Ce(III) sites with Sm(III) ions while maintaining the oxygen defect concentration. It was found that decreasing the concentration of Ce(III) by its replacement with Sm(III) results in decreasing the antioxidant (or radical scavenging) ability of nanoceria both in cell-free and cell containing experiments. This result highlights that the Ce(III) content is important for the ROS scavenging activity. While this result emphasizes the fact that Ce(III) ions are required for the transfer of electrons, it does not subdue the role of oxygen vacancies. Given the fact that the ROS scavenging activity of nanoceria at physiological pH is primarily dependent on Ce(III) concentration, the adsorbed superoxide radicals on oxygen vacancies cannot be reduced on their own and Sm(III) ions are incapable of participating in electron transfer with the adsorbed ROS species. Thus, this study proves that a minimum amount of Ce(III) is required for the antioxidant activity of nanoceria, but at the same time, it does not rule out the importance of oxygen vacancies. A study wherein nanoceria can be synthesized in its $Ce_2O_3$ form, with hexagonal crystal lattice, is required to prove that oxygen vacancies do not play a role. However, it is a challenging task especially without the use of any surface ligands and only a handful of studies have been reported with $Ce_2O_3$ synthesis ([121]).

Theoretical studies on the adsorption of oxygen molecules on nanoceria surfaces have clearly shown the importance of surface oxygen vacancies. The adsorption of oxygen is facilitated by the presence of oxygen vacancy on the ceria surface and the total energetics and electron transfer to and from oxygen will depend upon the location of the adsorbed oxygen from the vacancy site[58]. The energy barrier for converting superoxide to peroxide is merely 0.35 eV on the ceria surface, suggesting a facile conversion. The adsorption of oxygen and its conversion to superoxide or peroxide species is also facet-dependent[122]. At (111) facet, the adsorption of gas-phase oxygen is followed by its conversion to peroxide without any indication of the presence of superoxide, while at (100) facets, both superoxide and peroxide species can be found, although peroxides are more stable than the superoxide species. In addition, the presence of subsurface vacancies on nanoceria surfaces can play a role in enhancing its surface reactivity. The adsorption of surface oxygen promotes sub-surface oxygen vacancies towards the surface due to a concentration gradient[123]. However, such theoretical or experimental studies are conducted in pristine anhydrous environments while most of the biological reactions are carried out under an aqueous environment and in the presence of a large number of ions and dissolved organic matter, including proteins. Thus, a direct study comparing the interaction of various ions and biomolecules together with ROS and RNS is required to understand the overall importance of surface and sub-surface oxygen vacancies. In this context, it was recently demonstrated that the redox activities of nanoceria are pH dependent and that at pH 6, the ceria surface undergoes dynamic reconstruction in the presence of UV irradiation with partial removal of Ce(III) and associated surface and sub-surface vacancies. In contrast, at pH 9.0 the reconstruction of the surface allows the exposure of surface and sub-surface oxygen vacancies, underscoring the importance of vacancies in the redox activity of nanoceria[124].

### 5.2 Why oxidation state is important, and which oxidation state is important for pro-oxidant and antioxidant activity

The debate about the oxidation state stems from both the antioxidant and pro-oxidant abilities of nanoceria[125]. While the pro-oxidant activity (arising from oxidase and peroxidase activity) is required in a suite of sensing applications, it can have detrimental effects in vivo, as shown by the plethora of diseases associated with elevated ROS. In terms of oxidation state, the oxidase-like activity of nanoceria is directly related to the Ce(IV) oxidation state even though there is no systematic report on the oxidation state-dependent pro-oxidase activity[126]. However, peroxidase-like activity has been ascribed to a higher concentration of surface Ce(III)' highlighting that the decomposition of $H_2O_2$ may follow two different pathways, such as CAT-like pathway (i.e. direct conversion to oxygen) or peroxidase-like pathway (i.e. conversion to hydroxyl radicals)[73-74]. This is in direct contrast to the report from Pirmohamed et al., where nanoceria particles, rods and cubes were tested for CAT-like activity and showed a single pathway[48c]. It is also noted that many previous studies did not test the peroxidase-like activity of nanoceria, and the peroxidase pathway might be related exclusively to the presence of a right combination of surface Ce(III)/Ce(IV). Considering this and the anti-oxidative potential of nanoceria with a higher Ce(III)/Ce(IV) ratio, it is possible that the hydroxyl radicals do not desorb from the nanoceria surface and are immediately converted to oxygen and water. The interaction of nanoceria with a high concentration of $H_2O_2$ has been studied previously to some extent that seems to lean towards adsorption of peroxide on the nanoceria surface or the formation of a cerium-oxo-peroxo complex[59b]. The reaction pH, concentration of hydrogen peroxide, and surface coatings have a major effect on the activities of nanoceria that need to be explored further.

In contrast to these pro-oxidant activities, antioxidant activities in-vivo are usually dependent on surface Ce(III) concentration for preventing ROS and RNS mediated injury. None of the literature reports have ascribed the ROS protection of nanoceria to its surface Ce(IV) concentration, though some of the reports have hinted at the protection of nanoceria from $H_2O_2$-mediated oxidative injury by nanoceria with a lower Ce(III)/Ce(IV) ratio[35a, 77b]. Based on these arguments and other discussions throughout this review, it is safe to assume that for antioxidant related activities of nanoceria, it is important to have higher surface Ce(III)/Ce(IV) concentration, especially on occasions where the ROS injury pathway involves superoxide and hydroxyl radicals. It must be noted that during the ROS scavenging process, surface Ce(III) will be converted to surface Ce(IV) that will scavenge any intracellularly generated $H_2O_2$.

### 5.3 Is the activity biased by the disease models tested, or the observed activity is broad-spectrum?

The models tested so far have a bias towards ROS-mediated disease states and are largely produced via administration of exogenous agents[84, 88, 96], except for a few that have a genetic basis such as AMD or subarachnoid haemorrhage model of stroke[94, 104-106]. The models have largely been examined for the effects of nanoceria after the disease state has been established to ascertain their therapeutic benefit. However, models to establish radioprotection administer nanoceria prior to radiation[110-112]. Despite the limitations of these animal models to accurately represent the human conditions, the evidence suggests the ability of nanoceria to provide disease-modifying outcomes and thus worthy of further pre-clinical and, ultimately, clinical development[127].

### 5.4 Effect of surface coatings on redox activity

As theranostic nanomedicine, nanoceria must be protected against the adsorption of endogenous biomolecules such as proteins, lipids, vitamins, amino acids, nucleotides at their surface. Studies on the formation of a protein/chemical corona have shown that, following the adsorption, the particles tend to aggregate and ultimately sediment. Because the antioxidant properties of nanoceria are linked to their surface state, in particular to the density of oxygen vacancies and Ce(III) ions, any modification of this surface state via biomolecule adsorption or particle aggregation leads to a reduction in the catalytic properties[36c, 121a, 121c]. For nanoceria, more than for other particles, it is important to preserve the accessibility of free radicals to the reactive sites while ensuring stability in biological fluids through functionalization.

For functionalization, the coating can decrease the number of active sites and mitigation of catalytic properties. In this regard, a consensus has recently emerged on using the two-step post-synthesis technique, which combines simplicity, high yields of particle production, and leaves a potential for upscaling[79a, 79b]. In this approach, the particles and polymers are synthesized separately and later assembled following appropriate protocols, such as physical adsorption via electrostatic, H-bonding or hydrophobic interaction, or phase transfer of the coating agent between organic and aqueous solvents. This versatile technique takes advantage of an extensive library of functional polymers developed in recent years in polymer and coordination chemistry. Polymer coatings terminated with 2-10 kDa poly(ethylene glycol) macromolecules have shown excellent results in preserving catalytic properties of nanoceria. In comparison, techniques using ligands or low molecular weight molecules have shown mitigated results[79j]. Depending on the electronic properties and density of the adsorbed molecules, the antioxidant properties of nanoceria can be increased or decreased, with the ligands being generally insufficient to ensure colloidal stability.

**5.5 Is nanoceria stable in biological media?**

Studies have shown that when dispersed in synthetic or biological media, inorganic nanoparticles may undergo a certain number of transformations during their life cycle, i.e. from synthesis to applications. These transformations are associated with chemical reactions occurring at the nanoparticle surface with ions and molecules present in their close environments. These transformations mainly concern nanoparticle solubility, the structure of the crystalline cores and the surface state. With nanoceria, it is also important to focus on the oxidation state transformations, which are able to modify their catalytic activity.

Compared to other noble metal or metal oxide particles, nanoceria are relatively stable over time and display minor dissolution of surface atoms with aging. However, in certain conditions of pH, ionic strength, coating or light exposure, or a combination thereof, the particles are susceptible to dissolve over time and release Ce(III) and Ce(IV) ions into the medium. In low ionic strength aqueous buffer, Plakhova et al. established the extremely low solubility of 5 nm nanoceria ($\log K = -59.3 \pm 0.3$) by mass spectrometry and radioactive tracer methods and found that Ce(III) ions are the main species in the solution at $pH < 4$[128]. In the presence of phosphate ions (5 mM), Römer et al. have shown that the nanoceria surface and crystalline structure are altered upon aging (21 days), resulting in the formation of cerium phosphate ($CePO_4$) nanoparticles[129]. Nanoceria degradation was also observed for dispersions exposed to daylight for long periods (> weeks), including nanoceria coated with citrate ions[130] or polysaccharides[131]. In-plant growth medium, Schwabe et al. also found notice-able modifications of the nanoparticles when free ions (ferrous ion, phosphate) or a chelating agent (EDTA) was present in the surrounding medium, concluding that the nanoceria oxidation state plays a major role in the dissolution[132]. To sum up, these results show that nanoceria is not insoluble as often indicated and can release a significant amount of cerium, depending on the nanoparticle properties and solvent composition over a long period of time.

However, it should be kept in mind that as-synthesized nanoceria kept in their synthesis bath are, in general, stable for long-term without dissolution.

Besides dissolution and structural phase changes, the most common transformations encountered with engineered nanoparticles in biofluids concern their surface properties and the protein/chemical corona formation. Non-specific adsorption of ions, ligands or (macro) molecules is ubiquitous in nanotechnology, and it may induce a series of phenomena at the scale of the particles or of the dispersion. The protein/chemical corona can induce a series of interactions, including complexation reactions with the surface ions Ce(III) and Ce(IV) through electrostatic charge pairing or the building up of a barrier towards the active redox sites. These mechanisms are expected to lead to a decrease of the initial catalytic activity. Within a few exceptions[78, 79h], such effects have not been investigated in detail with nanoceria.
A second process concerns the colloidal stability of nanoceria. The stability of nanometer-sized nanoceria is controlled by interparticle repulsive interaction, generally mediated by electrostatic or steric potentials. Repulsive interaction potentials are required to compensate for the ubiquitous attractive van der Waals interaction. With a protein/chemical layer of a few nanometers at the surface, the effective inter-particle potential may become attractive, leading to aggregation, agglomeration or flocculation of the nanoceria. Such phenomena have been found in the recent literature for nanoceria coated with phosphate[29b, 48d] and citrate[79f] anions, polyelectrolyte[78] or PEGylated[79b, 79h, 133] brushes. This aggregation, in turn, modifies the nanoparticle properties such as size, morphology and diffusion. Aggregation also induces sedimentation, giving rise to concentration gradients and inhomogeneous dispersions. These factors can work together to decrease the redox properties of nanoceria and make them inappropriate for applications. Thus, it is important to control the functionalization of nanoceria for biological use, and that under appropriate coating conditions, it is possible to reconcile the colloidal stability and the preservation of native redox properties.

## 6.0 Future outlook – the road ahead

### *6.1 Factors overlooked in deciphering the mechanism of action of cerium oxide*

After more than a decade of research on the activity of nanoceria as either an antioxidant or prooxidant, a picture is evolving with a better understanding of its mechanistic action. Despite this, the reports are often focused on either the pro-oxidant activity or the antioxidant activity, but not both. Some of the irregularities in the reported literature can be ascribed to the difference in the synthesis of nanoparticles or the in vitro endpoint detection of pro or antioxidant activity. It is prudent to recognize that as an active redox system, the Ce(III)/Ce(IV) couple can show oxidative and reductive behaviour depending upon the redox properties of the interacting substrates and the pH of the medium. Thus, it is essential to understand its redox behaviour with biologically important redox couples and relevant biomolecules, such as, ROS or precursor molecules to the ROS/RNS generation. Two of the common pitfalls in deciphering the reaction mechanism of nanoceria are:

i. Both pro and antioxidant mechanisms are not mechanistically studied with the same set of nanoparticles with well-defined Ce(III)/Ce(IV) surface oxidation states. In this respect and as with any other nanomaterial, standardization of synthesis of well-characterised nanoceria and protocols for testing the enzymatic activities in cell-free and cell-containing media with the same set of well-characterised nanomaterials are important. Merrifield et al have compared the redox activity of $Ce_2O_3$ and $CeO_2$ of nearly equal sizes in presence of NOM and high concentration of ions. It was found that surface Ce(IV) in $CeO_2$ NPs transformed to Ce(III) while the surface Ce(III) in $Ce_2O_3$ remained unchanged in presence of high concentration of ions[121a]. More research looking at such particles will increase the understanding of the redox activity of nanoceria however, the role of ligands, in this case polymer, needs to be deconvoluted from the redox activity as different polymers may bind with different energy to the surface Ce(III) and Ce(IV) ions. Equally important is the fact that the estimation of Ce(III)/Ce(IV) ratio is generally interpreted from X-ray photoelectron spectroscopy (XPS) data, and being a surface-sensitive technique, XPS results are affected by surface contamination as well as X-ray induced reduction of Ce(IV) to Ce(III), and thus, any x-ray exposure of the nanoceria samples for analysis must be restricted to a very short time. Electron Energy Loss Spectroscopy (EELS) is another technique that can be useful for differentiating and quantifying the amount of Ce(III)/Ce(IV) ratio by comparing the ratio of M4 and M5 edge using different methods[121a, 121c]. Here again the electron beam damage should be taken into consideration before making any conclusive statements about the Ce(III)/Ce(IV) ratio.

ii. The in-vitro mechanism discussed in cell-free media is rarely correlated with the cell-containing media. Part of the reason for this is that the oxidative insult in cells may trigger the generation of ROS and RNS and the expression of other species that may be hard to distinguish. Thus, the mechanism of action in cell-based media for both pro- and anti-oxidant activities for a series of well-characterised nanoceria with different surface Ce(III)/Ce(IV) ratio should be conducted using specific probes for the identification of various redox species generated within the system. Such standardization cannot proceed without the quantification of nanoceria internalisation as the change in surface Ce(III)/Ce(IV) ratio will influence the size and surface charge, and hence the uptake will be different, resulting in variability.

Authors also note that the actual determination of redox potential of surface cerium ions on nanoceria has not been successfully reported thus far owing to difficulties in such measurements. Strategies to measure the redox potential of nanoparticles could pave the way for developing an understanding of its redox behaviour as a function of surface stoichiometry. This is an area of research that could benefit from the theoretical calculations.

***6.2 Ligand functionalized cerium oxide NPs – can ligands help in improving the selectivity or regeneration of the active oxidation state or cerium?***

Considering that the redox behavior of nanoceria is primarily dependent on the ability of surface cerium ions to participate in redox reactions with ROS and RNS species, it is possible to tune the redox behavior with surface ligands. This strategy essentially mimics the behaviour of metalloenzymes that show variable redox potentials based on the ligands attached to the metal centre. However, instead of doing this at a molecular level, this strategy will rely on this modification of a surface cerium ion bound to its lattice on one side and ligands on the other side. As discussed, surface modification of nanoceria with PEG and PAA coatings as well as phosphine-based ligands can modify their enzymatic activities[79a, 79j]. The activation of SOD activity of a nanoceria by pre-incubating with CuZn-SOD has also been reported wherein the electron transfer from the CuZn-SOD enhanced the SOD activity of nanoceria[134]. Such strategies can help to selectively tune the activity of nanoceria as only antioxidant or only pro-oxidant, increase or decrease the reactivity or promote faster regeneration by careful selection of the ligands. The high surface area of nanomaterials will ensure that enough active sites are still available for interaction with ROS and other intermediate species.

***6.3 Do we need cerium oxide lattice for the antioxidant activity or can cerium ions be incorporated in other structures for antioxidant activity?***

Single-atom nanozymes have emerged as a new class of materials that can perform the functions of the nanozymes with high selectivity and activity[135]. In these systems, active single atoms of reactive metals such as Cu, Fe, Co are immobilized on an inert substrate. A high distribution of these single-atom sites results in entering catalytic reactions with a high turnover ratio and high selectivity. In nanoparticle systems, the surface atoms participate in catalytic reactions. The physical structure and morphology of the nanoparticles has several heterogeneities in the form of surface terminations, edges or corners, all of which demonstrate different reactivity as well as selectivity towards a reaction. These surface heterogeneities also lead to several side reactions and hence, can be replaced by immobilizing the active metal centers within an organic (organometallic) structure[136]. The advantage of single atom nanozymes as compared to the conventional single-atom catalysis is that the nanozyme reactions are carried out at milder conditions. Thus, the single atoms can maintain a high dispersion as opposed to conventional catalysis, where the high-temperature catalysis results in diffusion-driven aggregation of single atoms.

The importance of the nanoceria lattice in imparting enzyme-like activity is unknown at present. If the oxygen vacancy plays no role in the adsorption of substrates on the surface, then the need for nanoceria lattice in depicting enzyme-like activity can be questioned. The enzyme-like activity shown by nanoceria can take advantage of single atom nanozyme chemistry by immobilization or entrapment of Ce(III)/Ce(IV) ions on substrates or embedded within an organometallic framework. The cerium ions can react with the ROS and RNS, while due to the absence of the lattice structure of ceria, the reactivity can be purely ascribed to the ions. It will also resolve the persistent debate about the role played by the fluorite lattice and oxygen vacancies in imparting nanoceria with its unique capability of ROS scavenging. Precise chemistry will have to be designed to preserve a single oxidation state of cerium and allow it to react with in-vitro generated ROS/RNS. It must be noted that free cerium ions have been shown to undergo Fenton chemistry with hydrogen peroxide[75]. Thus, the prevention of metal ions leaching from single metal (or ion) nanozyme is important to prevent metal ion-induced cytotoxicity. The immobilization of a single oxidation state of cerium ions may help to resolve its mechanism of action as it may prevent it from showing multiple types of enzyme activities.

**7.0 Conclusion**

Nanoceria is an important material with immense potential in modulating the generation or scavenging of ROS and RNS species. Several factors such as synthesis conditions, oxidation state, surface defects, type of the surface coating and pH of the reaction medium, influence its antioxidative or oxidative applications. These applications have widespread use in the diagnosis and therapy of various diseases caused by ROS and RNS and in other specific conditions such as the space medicine[137]. In view of the immense potential of nanoceria for scavenging ROS and RNS and mitigating diseases caused by oxidative stress, it is essential to resolve some of the long-standing issues and charter a new path forward for this exciting nanomaterial. Some of the recent discoveries have shown that a combination of two different nanozyme species can create a powerful antioxidant system that can replace the function of existing SOD enzymes in neuronal cells[138]. These hybrid materials will also be one of the paths forward for research on antioxidant nanomaterials. However, all new discoveries will need to be observed carefully for materials that show oxidation and reduction potential. One of the lessons for the nanozyme community from the complex antioxidant/pro-oxidant chemistries displayed by many nanozymes is to ensure that both the activities displayed by nanomaterials are tested before their wide-scale applications as antioxidant nanomaterials. Thankfully, in the case of nanoceria, the Ce(III)/Ce(IV) chemistry allows tuning its properties to display a range of activities that can be modulated exclusively for one type of activity.

**Author Biography**

**Megan Lord** is an Associate Professor in the Graduate School of Biomedical Engineering, UNSW Sydney, Australia. She obtained her Ph.D. from UNSW Sydney in 2006. Her research interests include developing materials and molecules that replicate the extracellular matrix for both the correct function of implantable medical devices and the controlled repair of tissues as a result of disease or injury. She has worked extensively on elucidating the internalization and antioxidant properties of nanoceria. She has received multiple awards for her academic and research achievements including NSW Young Tall Poppy Science Award in 2014. She is currently the president of the Matrix Biology Society of Australia and New Zealand and a member of many prestigious professional bodies.

**Jean François Berret** is a Professor at the University of Paris-Diderot, France and is a renowned expert in soft-condensed matter and biophysics. He is member of the Laboratoire Matière et Systèmes Complexes since 2005 and has a strong experience in fundamental and applied research. His research consists of developing novel functional nanostructures with stimuli-responsive features and biocompatibilityusing self assembly approaches. He has worked on the novel synthesis and application of nanoceria including development of functional coatings for biomedical applications.

**Sanjay Singh** is an Associate Professor at the Division of Biological and Life Sciences, Ahmedabad University, India. He completed his Ph.D. thesis from CSIR – National Chemical Laboratory, Pune, India. He worked as a postdoctoral fellow at the Burnett School of Biomedical Sciences, University of Central Florida, and at the Department of Pharmacology, College of Medicine, Pennsylvania State University, Pennsylvania, USA. He develops novel inorganic nanozymes exhibiting catalytic activities and uses these nanozymes for selective identification, imaging, and treatment of cancer cells. He also works on preparing novel nanoliposomal formulations for the efficient encapsulation and delivery of anticancer drugs and tumor suppressor genes to inhibit the proliferation of cancer cells. Dr. Singh has been awarded Endeavour Research Fellowship, Yamagiwa-Yoshida Memorial international award, and EMBO fellowship. He has been featured in Stanford's list of top 2% scientists in the world.

**Ajay Karakoti** is an Associate Professor at the University of Newcastle. He has predominantly worked on the applications of nanomaterials in healthcare, sensing and catalysis applications with a focus on the surface modification and its fundamental understanding that is critical to this project as well. He has contributed heaviliy to the field of enzyme mimicking properties of cerium oxide nanoparticles. His research focus includes understanding the mechanism of interaction of nanomaterials with rective oxygen species, modifying their redox activity with surface ligands and detailed surface and bulk characterization of materials. He has received multiple awards for his research including the NASI-Scopus Young Scientist Award in 2018 in India. He is a member of many professional bodies.

**Ajayan Vinu** is a Professor at the University of Newcastle and a leading researcher in the field of mesoporous materials. His contribution to the field of nanoporous materials is evident from the international ranking by 'Science Watch' as one of the top 15 researchers in the field. He introduced a new field of research on nanoporous nitrides and developed novel methods for making new nanoporous materials with different textural parameters and multiple functions. The impact of his research culminating in multiple reports of world's first mesoporous CN, boron nitrides, boroncarbon nitrides, biomolecules and fullerenes engineered through innovative polymeric chemistry and nanotemplating technologies for various applications in energy, environment and healthcare including adsorption of proteins, enzymes and other chemicals as well as their selective delivery. He has been ranked 6th in Australia among the Top 2% scientists in the field of "Materials" as per the database created by the Stanford University in 2020 reflecting his high-quality research and research impact that he made in this field. He is a fellow of many societies including Royal Society of Chemistry, Royal Australian Chemical Institute, World Academy of Ceramics, World Academy of Arts and Science, Asia Pacific Academy of Materials, and Maharastra Academy of Sciences.

TOC Graphic

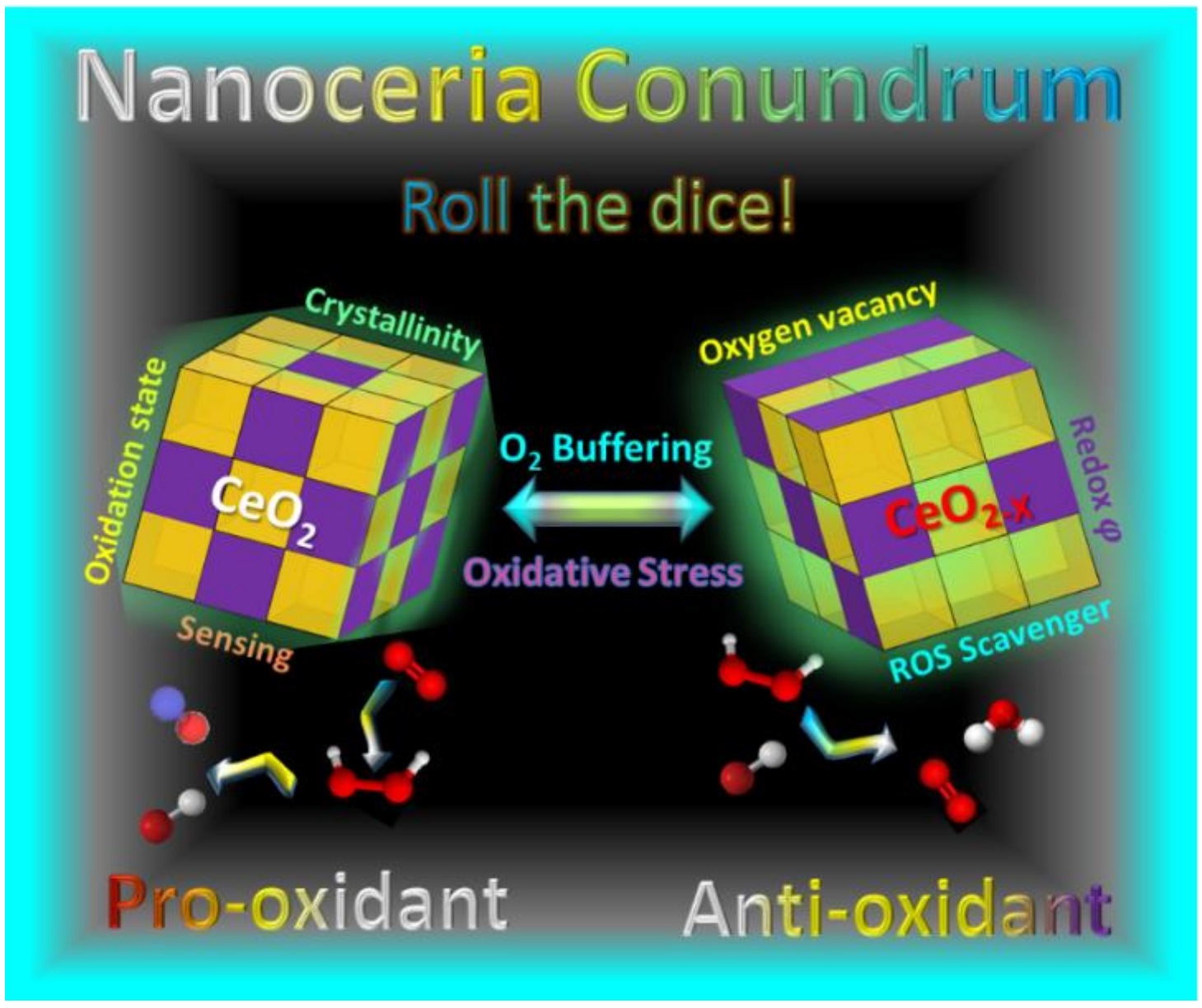

Cerium oxide nanoparticles (nanoceria) show unique oxidation state dependent enzyme mimicking activity. This review covers the current update on the proposed anti and pro-oxidant activity of nanoceria and the role of oxidation state, defects and surface coatings on its activity and biological applications. It summarizes current information and proposes the next agenda of research on nanoceria.